\documentclass[twocolumn,twocolappendix,trackchanges]{aastex7}

\usepackage{natbib}
\usepackage{comment}
\usepackage{amsmath}
\usepackage{algorithm}
\usepackage{amsfonts}
\usepackage{mathrsfs}
\usepackage{amssymb}
\usepackage{color}
\usepackage{graphicx} 
\let\tablenum\relax
\usepackage{siunitx}
\usepackage{placeins}

\newcommand{\cha}{{Chandra} }

\newcommand{\nus}{{NuSTAR} }
\newcommand{\xmm}{XMM-{Newton} }

\newcommand{\thei}{3--8\,keV }
\newcommand{\thtw}{3--24\,keV }
\newcommand{\eitw}{8--24\,keV }

\newcommand{\zetw}{0.5--2\,keV }
\newcommand{\twte}{2--10\,keV }

\newcommand{\no}{\nodata}
\newcommand{\DEL}[1]{}

\begin{document}

\defcitealias{Zhao2021}{Z21}
\defcitealias{Zhao2024}{Z24}
\defcitealias{Silver2026}{S26}

\correspondingauthor{Ross Silver}
\author[0000-0001-6564-0517]{Ross Silver}
\email{rosssilver.astro@gmail.com}
\affiliation{NASA Goddard Space Flight Center, Greenbelt, MD 20771, USA}
\affiliation{Southeastern Universities Research Association, Washington, DC 20005, USA}
\affiliation{School of Physics and Astronomy, University of Minnesota, Minneapolis, MN 55455, USA}

\author[0000-0002-2115-1137]{Francesca Civano} \email{francesca.m.civano@nasa.gov}
\affiliation{NASA Goddard Space Flight Center, Greenbelt, MD 20771, USA}

\author[0000-0002-7791-3671]{Xiurui Zhao} 
\email{xiurui.zhao.work@gmail.com}
\affiliation{Department of Astronomy, University of Illinois at Urbana-Champaign, Urbana, IL 61801, USA}
\affiliation{Cahill Center for Astrophysics, California Institute of Technology, 1216 East California Boulevard, Pasadena, CA 91125, USA}

\author[0000-0002-9041-7437]{Samantha Creech}
\email{s.creech@utah.edu}
\affiliation{Department of Physics \& Astronomy, University of Utah, 115 South 1400 East, Salt Lake City, UT 84112, USA}

\author[0000-0002-9895-5758]{S.\ P.\ Willner} \email{swillner@cfa.harvard.edu}
\affiliation{Center for Astrophysics \textbar\ Harvard \& Smithsonian, 
60 Garden Street, Cambridge, MA, 02138, USA}

\author[0000-0001-9262-9997]{Christopher N.\ A.\ Willmer} \email{cnawillmer@gmail.com}
\affiliation{Steward Observatory, University of Arizona,
933 N Cherry Ave, Tucson, AZ, 85721-0009, USA}

\author[0000-0001-8156-6281]{Rogier A.\ Windhorst} \email{Rogier.Windhorst@gmail.com}
\affiliation{School of Earth and Space Exploration, Arizona State University,
Tempe, AZ 85287-1404, USA}

\author[0000-0001-7592-7714]{Haojing Yan} 
\email{yanhaojing@gmail.com}
\affiliation{Department of Physics and Astronomy, University of Missouri,
Columbia, MO 65211, USA}

\author[0000-0002-6610-2048]{Anton M.\ Koekemoer}  \email{koekemoer@stsci.edu}
\affiliation{Space Telescope Science Institute,
3700 San Martin Drive, Baltimore, MD 21218, USA}

\author[0000-0003-3351-0878]{Rosalia O'Brien}
\email{robrien5@asu.edu}
\affiliation{School of Earth and Space Exploration, Arizona State University, Tempe, AZ 85287-1404, USA}

\author[0000-0002-6150-833X]{Rafael {Ortiz~III}}  \email{rortizii@asu.edu}
\affiliation{School of Earth and Space Exploration, Arizona State University,
Tempe, AZ 85287-1404, USA}

\author[0000-0003-1268-5230]{Rolf A.\ Jansen} \email{rolfjansen.work@gmail.com}
\affiliation{School of Earth and Space Exploration, Arizona State University,
Tempe, AZ 85287-1404, USA}

\author[orcid=0009-0007-0782-0721]{Gibson B.\ Bowling}
\email{gbbowlin@asu.edu}
\affiliation{School of Earth and Space Exploration, Arizona State University,
Tempe, AZ 85287-1404, USA}

\author[0000-0002-5319-6620]{Payaswini Saikia}
\email{payaswini.ssc@gmail.com}
\affiliation{Department of Astronomy, Yale University, PO Box 208101, New Haven, CT 06520-8101, USA}

\author[0000-0002-2203-7889]{W. Peter Maksym} \email{walter.p.maksym@nasa.gov}
\affiliation{NASA Marshall Space Flight Center, Huntsville, AL 35812, USA}

\author[0000-0002-1697-186X]{Nico Cappelluti} \email{ncappelluti@miami.edu}
\affiliation{Department of Physics, University of Miami, Coral Gables, FL 33124, USA}

\author[0000-0002-9286-9963]{Francesca Fornasini} \email{ffornasini@stonehill.edu}
\affiliation{Stonehill College, 320 Washington Street, Easton, MA 02357, USA}

\author[0000-0003-4223-7324]{Massimo Ricotti} \email{ricotti@umd.edu}
\affiliation{Department of Astronomy, University of Maryland, College Park, MD 20742, USA}

\author[0000-0001-6650-2853]{Timothy Carleton}  \email{tmcarlet@asu.edu}
\affiliation{School of Earth and Space Exploration, Arizona State University,
Tempe, AZ 85287-1404, USA}

\author[0000-0003-3329-1337]{Seth H.\ Cohen}  \email{seth.cohen@asu.edu}
\affiliation{School of Earth and Space Exploration, Arizona State University,
Tempe, AZ 85287-1404, USA}

\author[0000-0002-9984-4937]{Rachel Honor} 
\email{rchonor@asu.edu}
\affiliation{School of Earth and Space Exploration, Arizona State University, Tempe, AZ 85287-1404, USA}

\author[0000-0002-7265-7920]{Jake Summers} 
\email{jssumme1@asu.edu}
\affiliation{School of Earth and Space Exploration, Arizona State University,
Tempe, AZ 85287-1404, USA}

\author[0000-0002-9816-1931]{Jordan C.\ J.\ D'Silva}  \email{jordan.dsilva@research.uwa.edu.au}
\affiliation{International Centre for Radio Astronomy Research (ICRAR) and the
International Space Centre (ISC), The University of Western Australia, M468,
35 Stirling Highway, Crawley, WA 6009, Australia}
\affiliation{ARC Centre of Excellence for All Sky Astrophysics in 3 Dimensions
(ASTRO 3D), Australia}

\author[0000-0003-2714-0487]{Sibasish Laha} 
\email{sibasish.laha@nasa.gov}
\affiliation{Astrophysics Science Division, NASA Goddard Space Flight Center, Greenbelt, MD 20771, USA.}
\affiliation{Center for Space Science and Technology, University of Maryland Baltimore County, 1000 Hilltop Circle, Baltimore, MD 21250, USA.}
\affiliation{Center for Research and Exploration in Space Science and Technology, NASA/GSFC, Greenbelt, Maryland 20771, USA}

\author[0000-0001-7410-7669]{Dan Coe} 
\email{dcoe@stsci.edu}
\affiliation{Space Telescope Science Institute, 3700 San Martin Drive, Baltimore, MD 21218, USA}
\affiliation{Association of Universities for Research in Astronomy (AURA) for the European Space Agency (ESA), STScI, Baltimore, MD 21218, USA}
\affiliation{Center for Astrophysical Sciences, Department of Physics and Astronomy, The Johns Hopkins University, 3400 N Charles St. Baltimore, MD 21218, USA}

\author[0000-0003-1949-7638]{Christopher J.\ Conselice} \email{conselice@gmail.com}
\affiliation{Jodrell Bank Centre for Astrophysics, Alan Turing Building,
University of Manchester, Oxford Road, Manchester M13 9PL, UK}

\author[0000-0001-9065-3926]{Jose M. Diego}  \email{chemadiegor@gmail.com}
\affiliation{Instituto de F\'isica de Cantabria (CSIC-UC). Avenida. Los Castros
s/n. 39005 Santander, Spain}

\author[0000-0001-9491-7327]{Simon P.\ Driver} \email{Simon.Driver@icrar.org}
\affiliation{International Centre for Radio Astronomy Research (ICRAR) and the
International Space Centre (ISC), The University of Western Australia, M468,
35 Stirling Highway, Crawley, WA 6009, Australia}

\author[0000-0003-1625-8009]{Brenda Frye} 
\email{brendafrye@gmail.com}
\affiliation{Department of Astronomy/Steward Observatory, University of Arizona, 933 N Cherry Ave,
Tucson, AZ, 85721-0009, USA}

\author[0000-0001-9440-8872]{Norman A.\ Grogin}
\email{nagrogin@stsci.edu}
\affiliation{Space Telescope Science Institute,
3700 San Martin Drive, Baltimore, MD 21218, USA}

\author[0000-0001-6434-7845]{Madeline A.\ Marshall}  \email{madeline_marshall@outlook.com}
\affiliation{Los Alamos National Laboratory, Los Alamos, NM 87545, USA}

\author[0000-0003-3382-5941]{Nor Pirzkal} 
\email{npirzkal@stsci.edu}
\affiliation{Space Telescope Science Institute,
3700 San Martin Drive, Baltimore, MD 21218, USA}

\author[0000-0003-0429-3579]{Aaron Robotham} \email{aaron.robotham@uwa.edu.au}
\affiliation{International Centre for Radio Astronomy Research (ICRAR) and the
International Space Centre (ISC), The University of Western Australia, M468,
35 Stirling Highway, Crawley, WA 6009, Australia}

\author[0000-0003-0894-1588]{Russell E.\ Ryan, Jr.}  \email{rryan@stsci.edu}
\affiliation{Space Telescope Science Institute,
3700 San Martin Drive, Baltimore, MD 21218, USA}

\title{PEARLS: NuSTAR and XMM-Newton Extragalactic Survey of the JWST North Ecliptic Pole Time-Domain Field IV: X-Ray Variability Analysis}

\begin{abstract}
    \nus and \xmm have observed the {James Webb Space Telescope} (JWST) North Ecliptic Pole (NEP) Time-Domain Field (TDF) for almost five contiguous years starting in 2019. In that time, the NEP X-ray survey has accumulated 3.5\,Ms and 228\,ks of quasi-simultaneous \nus and \xmm observations, respectively. This paper presents variability results for the 112 \nus and 453 \xmm sources detected in this field, based solely on the X-ray photometric data. Four \nus sources and 11 \xmm sources varied in at least one band at $\ge$99\%  confidence. 
    The sources with redshift measurements show a relationship between luminosity and variability with 74\% of variable sources brighter than 5$\times$ the sensitivity limit of the survey. This is supported by about 1/3 of sources with more than 400 counts detected being variable and only 4 sources with fewer counts showing variability. Variability timescales are not well determined, but variability amplitude tends to be larger on longer timescales.
\end{abstract}

\section{Introduction}
Active galactic nuclei (AGN) are supermassive black holes (SMBHs) in the center of galaxies that are surrounded by an actively accreting disk of gas \citep{Soltan1982, Urry1995}. They are some of the most powerful non-transient emitters in the Universe and emit across the entire electromagnetic spectrum. In the last few decades, several works have found significant correlations between the mass of the SMBH and different host-galaxy properties such as the luminosity, mass of the bulge, and velocity dispersion \citep{Magorrian1998, Richstone1998, Gebhardt2000, Merritt2001, Ferrarese2005, Kormendy2013}. Each of these trends suggests that the SMBH and its host galaxy are connected and co-evolve \citep{Fiore2017, Martin-Navarro2018}. Therefore, extensive AGN studies are needed to understand how this connection affects galaxy evolution across cosmic time. 

One of the primary ways to study AGN is through their X-ray emission. A corona with hot, relativistic electrons interacts with photons from the accretion disk via inverse-Compton scattering and up-scatters these photons into the X-ray band \citep{Haardt1991}. This emission in the hard X-rays can be detected by the Nuclear Spectroscopic Telescope Array ({NuSTAR}), which is the first instrument capable of focusing hard X-rays \citep[3--79\,keV;][]{Harrison2013}. One of the great accomplishments of \nus has been its ability to resolve the cosmic X-ray background (CXB)---the diffuse X-ray emission that permeates the entire sky---around its peak of 20--40\,keV. Thanks to NuSTAR's sensitivity,  a factor of 10--100 better than coded-mask X-ray instruments,  \nus extragalactic surveys have resolved $\sim$35\% of the CXB at its peak energy \citep{Harrison2016}. Deep \nus surveys are one of the best tools for studying the AGN population across a large redshift range. 

The North Ecliptic Pole (NEP) Time-Domain Field  \citep[TDF;][]{Jansen2018} was chosen as a JWST target by the Prime Extragalactic Areas for Reionization Science (PEARLS) team for multiple reasons: 1) the field is within {JWST}'s northern continuous viewing zone (CVZ) and therefore can be observed year round; 2) there is low Galactic foreground extinction; and 3) there are no bright (AB $\leq$ 16 mag) foreground stars. The PEARLS team was guaranteed $\sim$47 hours of JWST time to observe during cycle 1 \citep[PI: R.\ Windhorst; program GTO-2738;][]{Windhorst2023}. In addition to the JWST data, the NEP-TDF has extensive coverage across the electromagnetic spectrum.\footnote{The full table is available at: \url{http://lambda.la.asu.edu/jwst/neptdf/}.}  

One of the best datasets in the NEP-TDF is the extensive quasi-simultaneous observations from \nus and the soft X-ray (0.2--12\,keV) instrument \xmm (XMM)\null. For almost five contiguous years starting in 2019, \nus and XMM have observed the NEP-TDF for 3.5\,Ms and 228\,ks, respectively. The results from the first 681\,ks taken during \nus cycle 5 were described by \citet[][hereafter \citetalias{Zhao2021}]{Zhao2021}, and the combined results from \nus cycles 5 and 6 (1561\,ks from \nus and 62\,ks from XMM) were published by \citet[][hereafter \citetalias{Zhao2024}]{Zhao2024}. The first two cycles detected 60 \nus sources and 286 \xmm sources, and  \citet{Creech2025} described spectral analysis of the 60 \nus sources. Subsequently, \citet[][hereafter \citetalias{Silver2026}]{Silver2026} found 75 \nus sources and 278 \xmm sources in the 2\,Ms of \nus data and 166\,ks of XMM data from \nus cycles 8 and 9. In total, 112 unique \nus sources and 453 unique XMM sources have been detected in the NEP-TDF, giving 504 unique sources altogether. 

Variability studies have been a common tool used to probe the
physical properties of AGN \citep{Ulrich1997, Peterson2004}, and the
main objective of the NEP-TDF \nus and \xmm observations was to
perform time-domain science. A simple light-travel time argument
gives first-order approximations of the sizes of the different
emitting regions. For example, reverberation mapping studies have
discovered the temperature profiles of accretion disks
\citep[e.g.,][]{Fausnaugh2016}, time lags between the continuum and
emission lines have been used to estimate the size of the broad-line
region \citep[BLR, e.g.,][]{Peterson2014}, and the inner size of the
dusty torus has been approximated by comparing delays between the
disk's visible-light emission and the near-infrared dust emission
\citep[e.g.,][]{Koshida2014}.

X-ray variability studies can be especially useful in studying AGN\null. The X-ray variability is typically larger in amplitude and on shorter timescales than lower-energy variability \citep{Ulrich1997, Peterson2004}, which suggests that the X-ray emission probes the innermost regions surrounding the SMBH\null. Moreover, X-rays have the ability to penetrate dust and gas more effectively than lower-energy emission, which enables variability studies of absorption in addition to emission \citep{Puccetti2014, Hernandez-Garcia2015, Torres-Alba2023, Cox2026}.

This paper analyzes the X-ray variability of all NEP-TDF X-ray-detected sources using all the  \nus and XMM observations. The paper is organized as follows: Section \ref{sec:data_reduc} briefly discusses the data reduction process for both the \nus and XMM observations. Section \ref{sec:var_method} describes the method to detect variability, and Section \ref{sec:results} reports the results. Section \ref{sec:disc} compares our results to other variability studies and looks for various trends in our results. Finally, Section \ref{sec:sum+concl} summarizes the paper and lists our conclusions. 
Where needed, this work assumes a \citet{Chabrier2003} IMF and a Planck 2018 cosmology with $ H_0 = 67.66$\,km\,s$^{-1}$\,Mpc$^{-1}$ and $\Omega_{\rm M} = 0.30966$ \citep{Planck2020}.

\section{Data Reduction} \label{sec:data_reduc}

\citetalias{Zhao2024} and \citetalias{Silver2026} provided in-depth discussion of how the cycles 5--9 data were reduced. Tables~\ref{Table:nus_cat_info} and~\ref{Table:xmm_cat_info} in Appendix~\ref{app_cat_desc} describe every column included in the entire 112-source \nus catalog and 453-source \xmm catalog published with this paper.  Table \ref{tab:obs_info} summarizes the observations.

\subsection{{NuSTAR}}
\begin{figure*}
    \centering
    \includegraphics[scale=0.75]{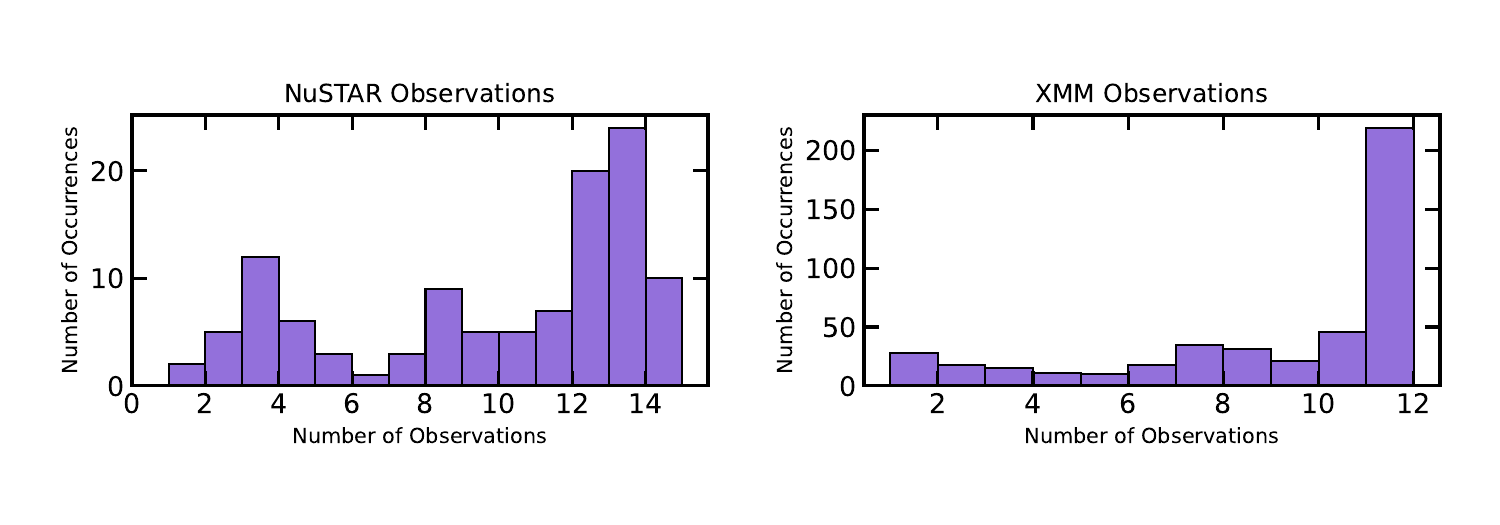}
    \vspace{-10ex}
    \caption{Histograms of the number of observations for each of the 112 \nus sources (left) and 453 XMM sources (right).}
    \label{fig:obs_hist}
\end{figure*}

The \nus NEP observations were taken during cycles 5 through 9, totaling 15 epochs and 47 individual observations spanning $>$4.5 years. The observations from \nus cycles 5 (PI: Civano, ID: 5192) and 6 (PI: Civano, ID: 6218)\footnote{This was a multi-cycle proposal, and observations also took place during cycle 7.} and the resulting catalogs were described by \citetalias{Zhao2021} and \citetalias{Zhao2024}. The observations from \nus cycles 8 (PI: Civano, ID: 8180) and 9 (PI: Civano, ID: 9267) were analyzed and published with the resulting catalog by \citetalias{Silver2026}. Figure \ref{fig:obs_hist} (left) shows the distribution of \nus observations for all 112 sources in our sample. 

The cycle 5 \nus data were reduced using HEASoft v6.27.2 and CALDB v.20200526; the cycle 6 data were reduced using HEASoft v6.29c and CALDB v.20211115; and the cycles 8 and 9 data were reduced using HEASoft v6.32 and CALDB v.20230718. For all observations, the level 1 raw data were calibrated, cleaned, and screened using the \texttt{nupipeline} script. 

For every observation, high count-rate intervals due to solar activity ($>$2 times the average count rate in the 3.5--9.5\,keV band full-field light curve) were filtered out. This is the band traditionally used to filter \nus surveys as it experiences the most radiation from solar flares. Cycle 8 experienced the most solar activity, with 13\% of the total exposure being removed, while cycles 5, 6, and 9 had only 1--4\% of the total exposure time removed. 

All vignetted exposure maps in the energy bands 3--8\,keV, 8--24\,keV, and 3--24\,keV were created using the NuSTARDAS tool \texttt{nuexpomap}. Background maps were created using the \texttt{nuskybgd}\footnote{\url{https://github.com/NuSTAR/nuskybgd}.} package \citep{Wik2014}. 
The 3--8\,keV band is similar to Chandra's energy range, while the 8--24\,keV band is much harder and thus able to capture emission from heavily obscured sources. The 3--24\,keV band maximizes NuSTAR's sensitivity.

\renewcommand*{\arraystretch}{1.2}
\begin{table*}
    \caption{\nus and \xmm Observations of the NEP-TDF from Cycles 5 through 9.}
    \centering
    \label{tab:obs_info}
    \vspace{-2ex}
    \begin{tabular}{ccccccc}
    \hline \hline
    \textbf{Cycle} & \textbf{Epoch} & \textbf{Date} & \textbf{NuSTAR IDs} & \textbf{Exp. (ks)} & \textbf{XMM ID} & \textbf{HST/JWST}\\
    \hline
    5 & 1 & 2019-10 & 60511001002, 60511002002, 60511003002 & 219.8 & \no & \no \\
      & 2 & 2020-01 & 60511004002, 60511005002, 60511006002 & 257.5 &  \no & \no \\
      & 3 & 2020-03 & 60511007002, 60511008001, 60511009001 & 203.7 & \no & \no \\
    \hline
    6 & 4 & 2020-10 & 60666001002, 60666002002, 60666003002 & 217.4 & 0870860101 & \no \\ 
      & 5 & 2021-01 & 60666004002, 60666005002, 60666006002 & 234.8 & 0870860201 & HST \\ 
    \hline
    7 & 6 & 2021-10 & 60666007002, 60666008002, 60666009002 & 180.7 & \tablenotemark{a} & HST \\ 
      & 7 & 2022-01 & 60666010002, 60666011002, 60666012002 & 246.7 & 0870860401 & HST \\ 
    \hline
    8 & 8 & 2022-08 & 60666013002, 60666014002, 60666015002 & 158.9 & 0870860501 & JWST \\
      & 9 & 2022-11 & 60810001002, 60810002002, 60810003002 & 296.1 & 0913590101 & JWST \\
      & 10 & 2023-02 & 60810004002, 60810005002, 60810006002 & 297.9 & 0913590501 & JWST \\
      & 11 & 2023-05 & 60810007002, 60810008002, 60810009002, 60810009004 & 189.7 & 0913590601 & JWST \\
    \hline
    9 & 12 & 2023-08 & 60910001002, 60910002002, 60910003002, 60910001004 & 238.8 & 0931420701 & \no \\
      & 13 & 2023-11 & 60910004002, 60910005002, 60910006002 & 238.9 & 0931420101 & \no \\
      & 14 & 2024-02 & 60910007002, 60910008002, 60910009002 & 222.6 & 0931420501 & \no \\
      & 15 & 2024-05 & 60910010002, 60910011002, 60910012002 & 193.8 & 0931420601 & \no \\
    \hline
    \end{tabular}
    \raggedright
    \tablecomments{The last column lists whether the observation had quasi-simultaneous observations with HST or JWST.}
    \tablenotetext{a}{The \xmm observation in 2021-10 (ID 0870860301) was wiped out by high particle background during the entire observation.}
\end{table*}

\subsection{XMM-{Newton}}
Simultaneous XMM-{Newton} observations were awarded for \nus cycles 6, 8, and 9 (see Table \ref{tab:obs_info} and Figure \ref{fig:obs_hist}, right). Observation 0870860301 from October 2021 was excluded from all analysis due to exceptionally high particle background during the entire observation. All \xmm data were reduced using the guidelines laid out by \cite{Brunner2008}, \cite{Cappelluti2009}, and \cite{Lamassa2016}. Descriptions of each package included in the \xmm Science Analysis System (SAS) can be found in the analysis threads.\footnote{\url{https://www.cosmos.esa.int/web/xmm-newton/sas-thread-src-find-stepbystep}} The cycle 6 data were reduced using SAS version 20.0.0 (\citetalias{Zhao2024}), and the cycles 8 and 9 data were reduced using SAS version 21.0.0 (\citetalias{Silver2026}). 

The SAS tasks \texttt{emproc} and \texttt{epproc} were used to generate the observational data files (ODF) from the three \xmm instruments (MOS1, MOS2, and PN). Any time intervals where the count rate exceeded 0.2 (MOS) and 0.3 (PN) counts s$^{-1}$ were excluded. The clean event files were then used to generate the MOS1, MOS2, and PN images in the \zetw and \twte bands. The tools \texttt{eexpmap} and \texttt{esplinemap} were used to generate the exposure maps and background maps, respectively.

\section{Variability Analysis Methodology} \label{sec:var_method}

The primary purpose of the X-ray observations of the NEP-TDF is to study source variability. \citetalias{Zhao2024} presented the first \nus contiguous survey to study hard X-ray variability up to 24\,keV. Adding two subsequent cycles of data has given 15 epochs of \nus observations spanning $>$4.5 years and 11 epochs of \xmm observations  to analyze for X-ray variability.  

We followed the same variability analysis established by \citetalias{Zhao2024}. The traditional X-ray variability method from \cite{Yang2016} is not applicable to our sources because they do not possess the required high counting statistics compared to background levels. For example, the median net source counts of the 
\nus cycles 8+9 sources in the \thtw band is 110 counts, while the median background counts is 350. When broken down by individual epoch, this will produce a low S/N and an unreliable net count rate. Because of this, \citetalias{Zhao2024} developed a Bayesian pipeline to analyze variability in this low-count regime. 

\begin{figure}
    \centering
    \includegraphics[width=1\linewidth]{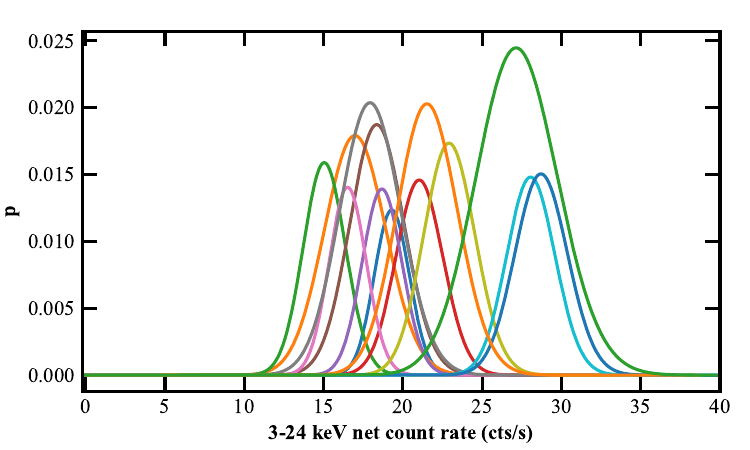}
    \caption{The \thtw net count-rate PPD for NuS59 9, the brightest and most variable source in our sample. Each color represents a different epoch.}
    \label{fig:ppd_blazar}
\end{figure}

The \citetalias{Zhao2024} pipeline utilizes the Bayesian methodology developed by \cite{Primini2014} and first used in the Chandra Source Catalog\footnote{\url{https://cxc.cfa.harvard.edu/csc/}.} \citep[CSC;][]{Evans2024}. This method calculates the probability distribution of the expected net source counts from every epoch. This approach works in low-count regimes because it implements Poisson statistics. Equation (16) from \cite{Primini2014} was used to calculate the net count posterior probability distribution (PPD) with noninformative prior distributions. A 20$\arcsec$ circle was used to extract the total and background counts from each epoch. Considering that not all sources were detected at every epoch, we used a fixed source position to extract the counts from every epoch in all five energy bands. Then we calculated the net count rates using the effective exposure from each observation. Figure \ref{fig:ppd_blazar} shows the \thtw net count-rate PPD for the brightest and most variable source in our sample, NuS59 9. 

For each source, we calculated a probability that it was variable using the $\chi^2$ test from the CSC method. This method calculates the difference between the most probable flux from each epoch compared with the source's most probable flux from the survey as a whole. The probability $p$ of a false positive was calculated using Equation (10) of Nowak 2016.\footnote{\url{https://cxc.cfa.harvard.edu/csc/memos/files/Nowak_csc2_variability_and_color.pdf}.} This means that the smaller the value of $p$, the higher the likelihood that the source is variable in that energy band.

\section{Results} \label{sec:results}

\subsection{{NuSTAR}}

\begin{figure}
    \centering
    \includegraphics[width=1\linewidth]{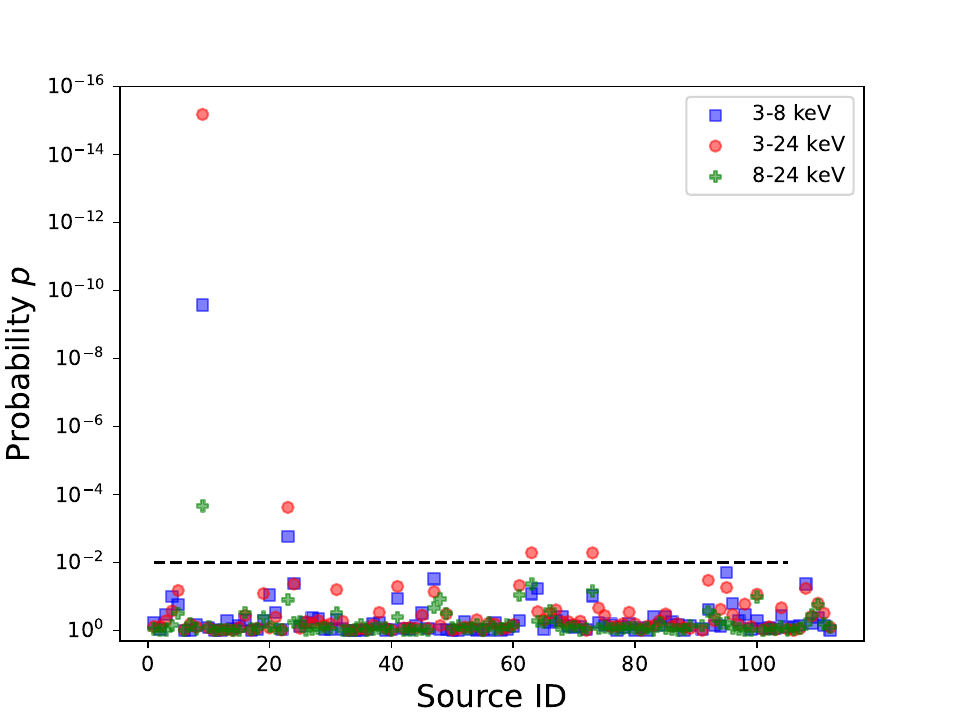}
    \caption{Variability probability $p$ for all 112 \nus sources detected in either cycles 5+6 or 8+9. The three energy bands shown are \thei (blue squares), \thtw (red circles), and \eitw (green diamonds). The dotted black line represents a 1\% false positive rate.}
    \label{fig:nus_var_p_no_blaz}
\end{figure}

Figure \ref{fig:nus_var_p_no_blaz} displays the $p$ values in the bands 3--8\,keV, 3--24\,keV, and 8--24\,keV for all 112 sources from the cycles 5+6 and 8+9 \nus surveys.  We use $p = 0.01$ as a threshold to identify variable sources. This threshold was chosen so that only $\sim$1 out of the 112 NuSTAR-detected sources is likely to be a false positive. In total, four sources were found to be variable in at least one energy band. Of these four, two were variable only in the \thtw band (NuS59 63, 73), one was variable in the \thei and \thtw bands (NuS59 23), and NuS59 9, the brightest source in this sample, was the only source with significant variability in all three bands. The level of variability for each source in all three bands is in Table~\ref{tab:nus_var10}. 

The preliminary variability analysis done by \citetalias{Zhao2024} found three out of the 65 sources to be variable with $p\lesssim 0.01$. Only two of these sources still display variability at this threshold after including the cycles 8+9 observations: NuS59 9 and NuS59 23. However, NuS59 19 was above the $p\leq 0.01$ threshold when including all 15 epochs ($p = 0.0818$ in the 3--24\,keV band). It is not surprising that some sources would change from below the $p\leq 0.01$ limit ($p = 0.0053$) to above it. When more observations were added into the sample, the most probable flux from the survey as a whole changed. In the case of NuS59 19, the most probable flux decreased, thus becoming more similar to the lowest flux observation and decreasing the overall probability of variability (Fig.~\ref{fig:lc_nus_19} in the Appendix).

\renewcommand*{\arraystretch}{1.2}
\begin{table*}
    \caption{Properties of the 4 Variable \nus Sources}
    \centering
    \hspace{-0.8in}
    \label{tab:nus_var10}
    \begin{tabular}{cccccccccr}
    \hline \hline
    \textbf{NuS59} & \textbf{XMM69} & \textbf{R.A.} & \textbf{Decl.} & \boldmath{$z$} & \boldmath{$p_{3-24\,keV}$} & \boldmath{$p_{3-8\,keV}$} & \boldmath{$p_{8-24\,keV}$} & \textbf{Num.\ Obs.}&\textbf{Counts}\\
    \hline
    9 & 299 & 260.8072 & 65.7975 & 1.4411 & $7 \times 10^{-16}$ & $3 \times 10^{-10}$ & 0.0002 & 13&4412 \\
    23 & 222 & 260.6730 & 65.7127 & 0.1791 & 0.0002 & 0.0017 & 0.1230 & 12 &1526 \\
    63 & 131 & 260.4659 & 66.0817 & 0.913 & 0.0052 & 0.0832 & 0.0432 & 2 & 181 \\ 
    73  & \no & 261.2084 & 65.9242 & \no & 0.0052 & 0.0952 & 0.0693 & 5 &74 \\
    \hline
    \end{tabular}
\raggedright
\tablecomments{NuS59 is the ID number in the combined Cycles~5--9 \nus catalog (Table~\ref{Table:nus_cat_info}), and XMM69 is the ID in the combined Cycles~6--9 \xmm catalog (Table~\ref{Table:xmm_cat_info}).  Right ascension and declination are J2000 \nus coordinates in decimal degrees.  
Counts are the sum of 3--24\,keV net counts in cycles 5--9.}
\end{table*}

\subsection{XMM-{Newton}}

\begin{figure}
    \centering
    \includegraphics[width=1\linewidth]{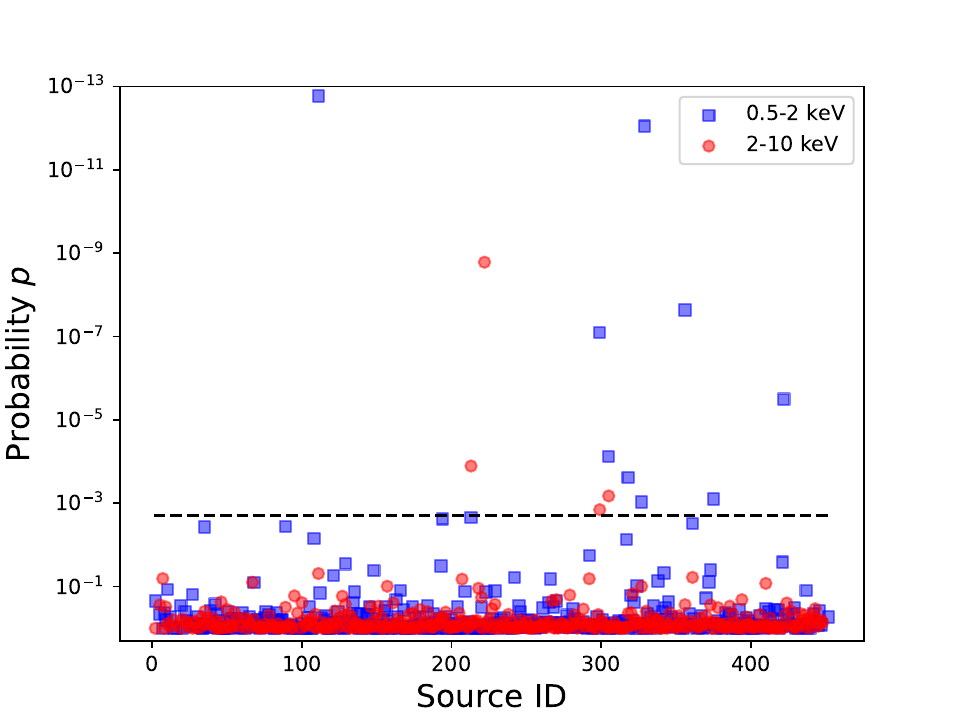}
    \caption{Variability probability $p$  for all 453 \xmm sources detected in either cycles 6 or 8+9. The two energy bands shown are \zetw (blue squares) and \twte (red circles). The dotted black line at $p = 0.002$  represents the level for a single false-positive detection in the sample.} 
    \label{fig:xmm_var_prob_p}
\end{figure}

Figure \ref{fig:xmm_var_prob_p} displays  $p$ for all 453 \xmm sources in the \zetw and \twte bands. Of the 453, 11 sources were variable with $p\leq 0.002$. Similarly to the \nus procedure, this threshold was selected so that only $\sim$1 target out of the 453 XMM detected sources is likely a false positive. Seven sources were variable solely in the \zetw band, two sources were variable only in the \twte band, and two sources were variable in both bands. The $p$-values for these sources can be seen in Table \ref{tab:xmm_var24}. 
Because \citetalias{Zhao2024} analyzed only one cycle with XMM observations, there are no previous variability results to compare with.

\renewcommand*{\arraystretch}{1.1}
\begin{table*}
\centering
    \caption{Properties of the 11 Variable \xmm Sources}
    \label{tab:xmm_var24}
\hspace*{-0.6in}
    \begin{tabular}{ccccccccr}
    \hline \hline
    \textbf{XMM69} & \textbf{NuS59} & \textbf{R.A.} & \textbf{Decl.} & \boldmath{$z$} & \boldmath{$p_{0.5-2\,keV}$} & \boldmath{$p_{2-10\,keV}$} & \textbf{Num.\ Obs.}& \textbf{Counts}\\
    \hline
    111 & 21 & 260.4320 & 65.7676 & 0.7813 & 2 $\times$ 10$^{-13}$ & 0.0487 & 11 & 3252 \\
    213 & 16 & 260.6399 & 65.8449 & 1.3369 & 0.0022 & 0.0001 & 11 & 1796 \\ 
    222 & 23 & 260.6717 & 65.7114 & 0.1791 &  0.9970 & 2 $\times$ 10$^{-9}$ & 11 & 1019 \\
    299 & 9 & 260.8091 & 65.7961 & 1.4411 & 8 $\times$ 10$^{-8}$ & 0.0014 & 11 & 10699\\
    305 & 5 & 260.8241 & 65.9435 & 0.4949 & 8 $\times$ 10$^{-5}$ & 0.0007 & 11 & 937\\
    318 & 4 & 260.8671 & 65.9873 & 1.4252 & 0.0002 & 0.4410 & 9 & 664\\
    327 & 20 & 260.8972 & 65.8175 & 0.7791 & 0.0009 & 0.0991 & 11 & 534\\
    329 & \no & 260.9062 & 65.6507 & \no & 9 $\times$ 10$^{-13}$ & 0.9930 & 8 & 180\\
    356 & 43 & 260.9653 & 65.8348 & 0.6726 & 2 $\times$ 10$^{-8}$ & 0.2720 & 11& 673 \\
    375 & 1 & 261.0066 & 66.0096 & 0.8918 & 0.0008 & 0.7920 & 7 & 1770\\
    422 & \no & 261.1716 & 65.9465 & 0.498 & 3 $\times$ 10$^{-6}$ & 0.9540 & 7 & 196\\
    \hline
    \end{tabular}
\raggedright
\tablecomments{NuS59 is the ID number in the combined Cycles~5--9 \nus catalog (Table~\ref{Table:nus_cat_info}), and XMM69 is the ID in the combined Cycles~6--9 \xmm catalog (Table~\ref{Table:xmm_cat_info}).  Right ascension and declination are J2000 \xmm coordinates in decimal degrees.  Counts are the sum of all 0.5--10\,keV net counts in cycles 6--9.}
\end{table*}

\section{Discussion} \label{sec:disc}

\subsection{Multiwavelength Campaign}

\subsubsection{Matching Procedure}
Our procedure for matching X-ray sources to multi-wavelength catalogs utilized the maximum-likelihood estimator \citep[MLE;][]{Sutherland1992}, which uses both the physical properties of the source and positional separation to determine the most likely counterpart. The 1) flux and offset for each candidate counterpart as well as  2) the positional uncertainties and flux distribution of the entire survey were used to calculate the likelihood ratio
\begin{equation}
    LR = \frac{q(m)f(r)}{n(m)},
\end{equation}
where $m$ is the catalog magnitude of the potential counterpart, ${n(m)}$ is the magnitude distribution of background sources near the target source, ${q(m)}$ is the expected magnitude distribution of the true multiwavelength counterpart, $r$ is the positional separation between the potential counterpart and the target source, and  $f(r)$ is the probability distribution of positional uncertainties. These LR values were then compared with the LR threshold (LR$_{\rm th}$), which was determined by the reliability ($R$) and completeness ($C$) of the X-ray survey. In this work, LR$_{\rm th}$ was chosen to maximize $(R + C) / 2$. The potential counterparts are described as 1) \textit{secure} if only one longer-wavelength counterpart has $\rm LR > LR_{\rm th}$, 2) \textit{ambiguous} if multiple longer-wavelength counterparts have $\rm LR > LR_{\rm th}$, or 3) \textit{unidentified} if the X-ray source has no potential longer-wavelength counterparts with $\rm LR > LR_{\rm th}$ within the search radius of 5\arcsec.

\subsubsection{HST}
As described by \citetalias{Zhao2024} and \citetalias{Silver2026}, the NEP-TDF also has significant HST and JWST coverage. As a part of the ``TREASUREHUNT'' program (HST GO Program ID 16252, Jansen \& Grogin), the HST Wide Field Camera on the Advanced Camera for Surveys covered about 88\,arcmin$^2$ in the central area of the \xmm and \nus coverage \citepalias[][Fig.~13]{Silver2026}.  These observations, with the F275W, F435W, and F606W filters, occurred between 2017 and 2022, creating the potential for time-domain science. \cite{Obrien2024} analyzed these data and found 12 transients and 190 variable candidates. \citeauthor{Obrien2024} attributed most of the transients to supernovae and most of the variable sources to AGN. 

Our positional crossmatching of the HST variables to the X-ray catalogs found two of the 190 HST variables to lie within 1.5\arcsec \space of an \xmm source. All other HST variables are at least 4.9\arcsec \space  from the nearest \xmm source, making them unlikely matches.
The source with the closest match is XMM69 327 = NuS59 20 = HST\_ID 36152 = V174. This source is identified as an AGN at $z$ = 0.7791 \citep{Obrien2024} and is an X-ray variable (Table~\ref{tab:xmm_var24}). 
The other good match is XMM69 137 = HST\_ID 42426 = V009. This source was not detected by {NuSTAR}, has unknown redshift, and showed no signs of variability in the \xmm data ($p = 0.446$ in the \zetw band and $p = 0.861$ in the \twte band).
The small number of matches is unsurprising considering how few of the HST sources, either variables or non-variables, are detected in X-rays. \\
\indent \cite{Obrien2024} crossmatched the HST transients to published and preliminary X-ray catalogs and found three possible X-ray matches for their 12 transient sources. We confirm one of them, XMM69 291 = HST T7 \citep[][their Table~2]{Obrien2024}.  This source has unknown redshift and shows no evidence of \xmm variability ($p = 0.966$ in the \zetw band and $p = 0.760$ in the \twte band).  The second transient \cite{Obrien2024} identified with an X-ray source, HST T10, is 5.6\arcsec \space  from XMM69 132 (unknown $z$, not variable) and therefore probably unrelated. \cite{Obrien2024} considered HST T11 a likely match to NuS59 12, but that \nus source is more likely a match to XMM69 160, which is $>$30\arcsec\ from the transient and therefore unrelated.

\subsubsection{JWST}
The existing JWST observations have little area observed more than once, and no variability study has been performed to date. However, JWST analyses have identified counterparts of our sample of variable X-ray sources. For example, \cite{Ortiz2024} identified 66 Seyfert galaxies and weak AGN in the NEP-TDF. Of these 66, four are variable XMM sources (XMM69 213/222/327/356), and one is also a variable \nus source (NuS59 23). Additionally, \cite{Willner2026} identified 208 JWST counterparts (206 NIRCam, 2 NIRISS-only) of $S(3 \rm GHz) > 5$\,$\mu$Jy radio sources. Of these, three are variable in the XMM data (XMM69 213/222/327). All three sources are also variable in 
\nus (NuS59  16/23/20).

\subsection{Host-Galaxy Properties}
\label{s:hosts}
\begin{figure}
    \centering
    \includegraphics[width=1.2\linewidth]{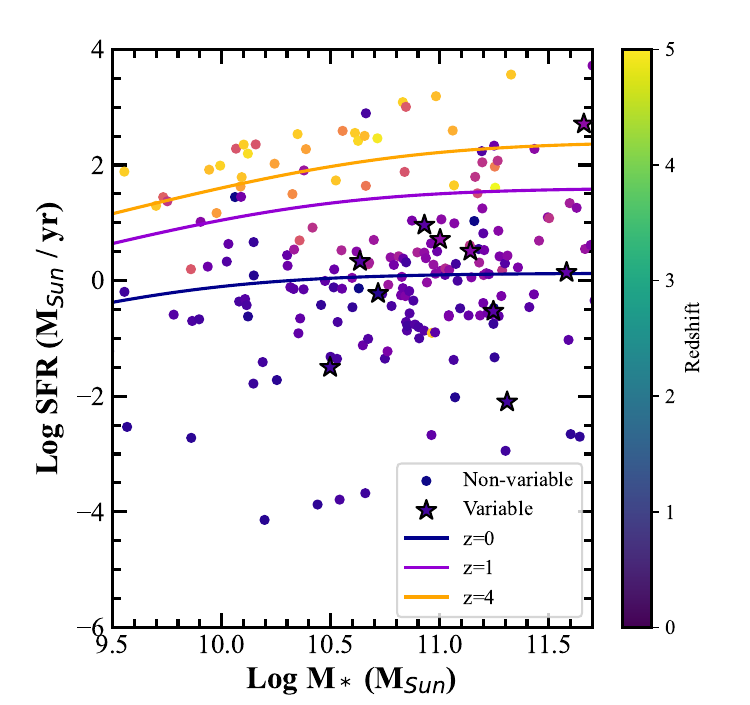}
    \caption{ SFR versus  $M_*$ for the 258 non-variable sources (circles) and 10 variable sources (black-outlined stars). The color map represents the redshift range of the entire sample. The curves represent the SFMS from \cite{Popesso2023} for $z=0$, $z=1$ (the median $z$ of our sample), and $z=4$ (the maximum spec-$z$ of our sample).}
    \label{fig:SFR_vs_M-S}
\end{figure}

R. Ortiz et al.\ (ApJ, submitted, 2026) modeled the radio to X-ray spectral energy distributions of the X-ray host galaxies with CIGALE \citep{Boquien2019}.  Of the 453 \xmm sources, 261
have unambiguous counterparts with enough long-wavelength data to be modeled.
Figure~\ref{fig:SFR_vs_M-S} shows SFR plotted against  $M_*$  for both samples. All 10 of the variable sources that can be modeled (the ones with redshifts shown in Table~\ref{tab:xmm_var24}) reside in galaxies with $\log M_*\ge10.5$ $M_{\odot}$. These galaxies' median $\langle\log M_*\rangle=11.07$ compared to $\langle\log M_*\rangle=10.67$ for all galaxies that can be modeled. Part of the difference is that variability can be detected only in bright X-ray sources, which may tend to reside in high-mass hosts. If the sample is limited to hosts with $>$400 \xmm counts (Table~\ref{tab:xmm_var24}), $\langle\log M_*\rangle=10.85$, but 8 of the 26 galaxies with that many counts are variable.  If those are removed, the rest have $\langle\log M_*\rangle=10.44$, comparable to $\langle\log M_*\rangle$ for the whole sample. The two conclusions are that $\sim$1/3 of galaxies with sufficient counts to detect modest variability are in fact variable, and those sources tend to reside in higher-mass hosts than just-as-bright non-variable (or at least non-detected as variable) X-ray sources. In contrast, SFR and the AGN fraction determined by CIGALE $f_{\rm AGN}$ show little difference between the non-variable and variable populations, but the sample size is small.

\subsection{Luminosity vs Redshift}
\begin{figure}
    \centering
    \includegraphics[width=1\linewidth]{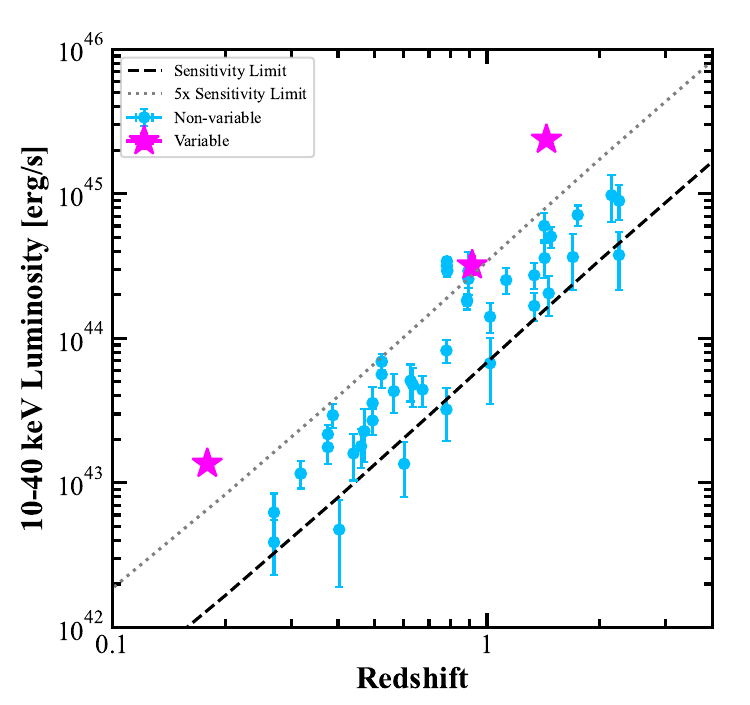}
    \caption{\nus 10--40\,keV luminosity versus redshift. Pink stars represent the three variable sources, and blue circles represent the 43 non-variable sources. The plotted luminosities are the average over all observations. The black dashed line shows the sensitivity limit of this survey, and the gray dotted line is 5$\times$  the sensitivity limit. Error bars are present for the three variable sources, but they are smaller than the magenta stars.}
    \label{fig:nus_lum_vari}
\end{figure}

Three of the 4 variable \nus sources have confirmed redshifts (Table \ref{tab:nus_var10}).\footnote{NuS59  73 does not have a soft X-ray counterpart, and the \nus positional uncertainty is too large to find a reliable visible--IR counterpart and thus a redshift.} Figure \ref{fig:nus_lum_vari} compares the 10--40\,keV luminosities of these three sources with the 43 non-variable \nus sources with known redshifts. Variability and luminosity are connected in that all of the variable sources have luminosities greater than 5$\times$ the sensitivity of the survey, and 3 of 7 sources (43\%) above this limit exhibit variability.  In terms of counts, 2 of 9 sources with $>$400 counts exhibit variability, comparable to the $\sim$1/3 found for \xmm (Section~\ref{s:hosts}).  Selection is a factor, but some distant, luminous \nus sources have large enough amplitudes to be detected as variable even with few counts.

\begin{figure*}
    \centering
    \includegraphics[width=0.6\linewidth]{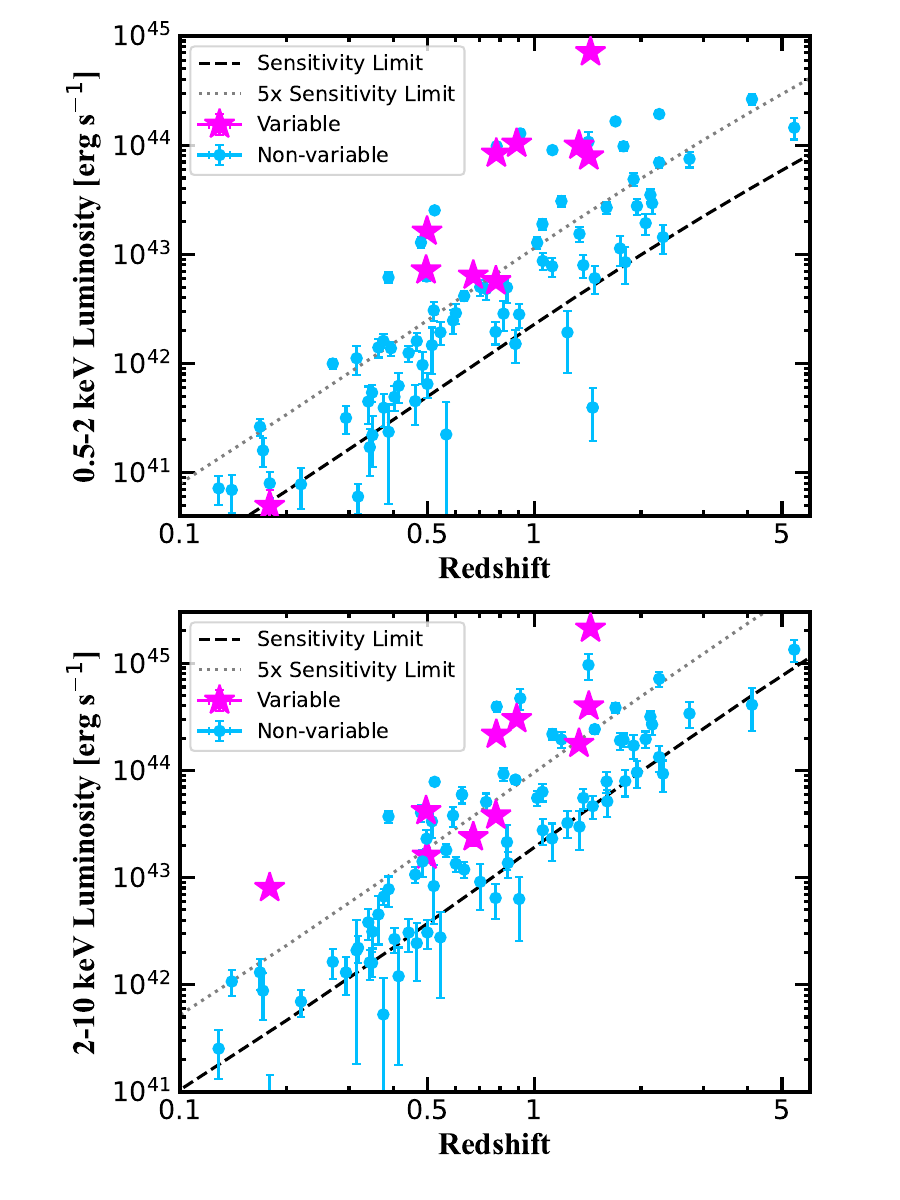}
    \caption{\xmm \zetw and \twte luminosities versus redshift. Pink stars represent the 10 variable sources and blue circles represent the 81 non-variable sources. Luminosities are the average over all observations. The black dashed line shows the sensitivity limit of this survey, and the gray dotted line shows 5$\times$ the sensitivity limit. Error bars are present for the 10 variable sources, but they are smaller than the magenta stars.}
    \label{fig:xmm_lum_vari}
\end{figure*}

Redshifts are known for 10 out of the 11 variable XMM sources (Table~\ref{tab:xmm_var24}). 
Those sources' average \zetw and \twte luminosities are plotted against redshift in Figure \ref{fig:xmm_lum_vari}. Similarly to the \nus sources, the \xmm sources show a connection between  variability and luminosity. In the \zetw band, 8/10 (80\%) of the variable sources are above 5$\times$ the sensitivity of the survey, and 8/32 (25\%) of all sources above this threshold are variable. Meanwhile, only 2/59 (3\%) of sources below the 5$\times$ limit are variable. 
This behavior also exists in the \twte band but is not quite as strong. 6/10 (60\%) of the variable sources are brighter than 5$\times$ the survey limit, and 6/25 (24\%) of all sources above this limit are variable. Meanwhile, only 4/66 (6\%) below the 5$\times$ limit are variable. 

In the combined (\nus + XMM) sample,  17/23 (74\%) of the sources have counts 5$\times$ the respective survey limit, and 17/64 (27\%) of sources brighter than 5$\times$ the survey limit exhibit variability.  Only 
6/161 (4\%) of sources below this limit exhibit variability. 
However, this result has potential biases. Variability will always be easier to detect in sources that have smaller photometric uncertainties, i.e., are brighter.  Second, the variability analysis includes only observations with confirmed detections; i.e., any observation that produced only an upper limit was excluded. Brighter sources will have more confirmed detections and thus more opportunities to exhibit variability. These biases are somewhat but not entirely mitigated by considering total counts as done above. \\
\indent Combining the \nus and XMM results, as done in this section for  simplicity, produces a further inaccuracy. The stricter cutoff ($p\leq 0.002$) for the XMM sources tends to select sources with stronger variability, larger X-ray luminosities, and thus smaller photometric uncertainties. Comparison between the \nus and XMM results is mainly limited by the small numbers of variables detected, but the different selection criteria also contribute. However, the combined samples still provide better statistics than either sample alone.

\subsection{Variability Timescales} \label{sec:var_time}
The time between observations affects the level of variability a source will show. We compared the fluxes (with uncertainties) from every observation pair and calculated the corresponding $\chi^2$ assuming no variability. The equation used is:
\begin{equation} \label{eq:90conf_lev}
    \chi^2 = \frac{(F_1-\mu)^2}{(\delta F_1)^2} + \frac{(F_2-\mu)^2}{(\delta F_2)^2},
\end{equation}
\noindent where $F$ is the flux for each observation, $\delta F$ is the uncertainty of each flux measurement, and $\mu$ is the mean of the two fluxes for each observation pair. For an object with constant flux, 90\% of the time $\chi^2$ will be less than 2.706 and 99\% of the time $<$6.635.  Larger $\chi^2$ suggests variability on the timescale corresponding to the interval between the observations.  

The first \nus observation analyzed in this work occurred in cycle 5 during 2019 October, and the last  occurred in cycle 9 during 2024 May. Therefore, the maximum time between observations is 4.5 years. Meanwhile, the shortest time between two observations was two months (epochs 2 and 3 from cycle 5). 

\begin{figure*}
    \centering
    \includegraphics[width=1.0\linewidth]{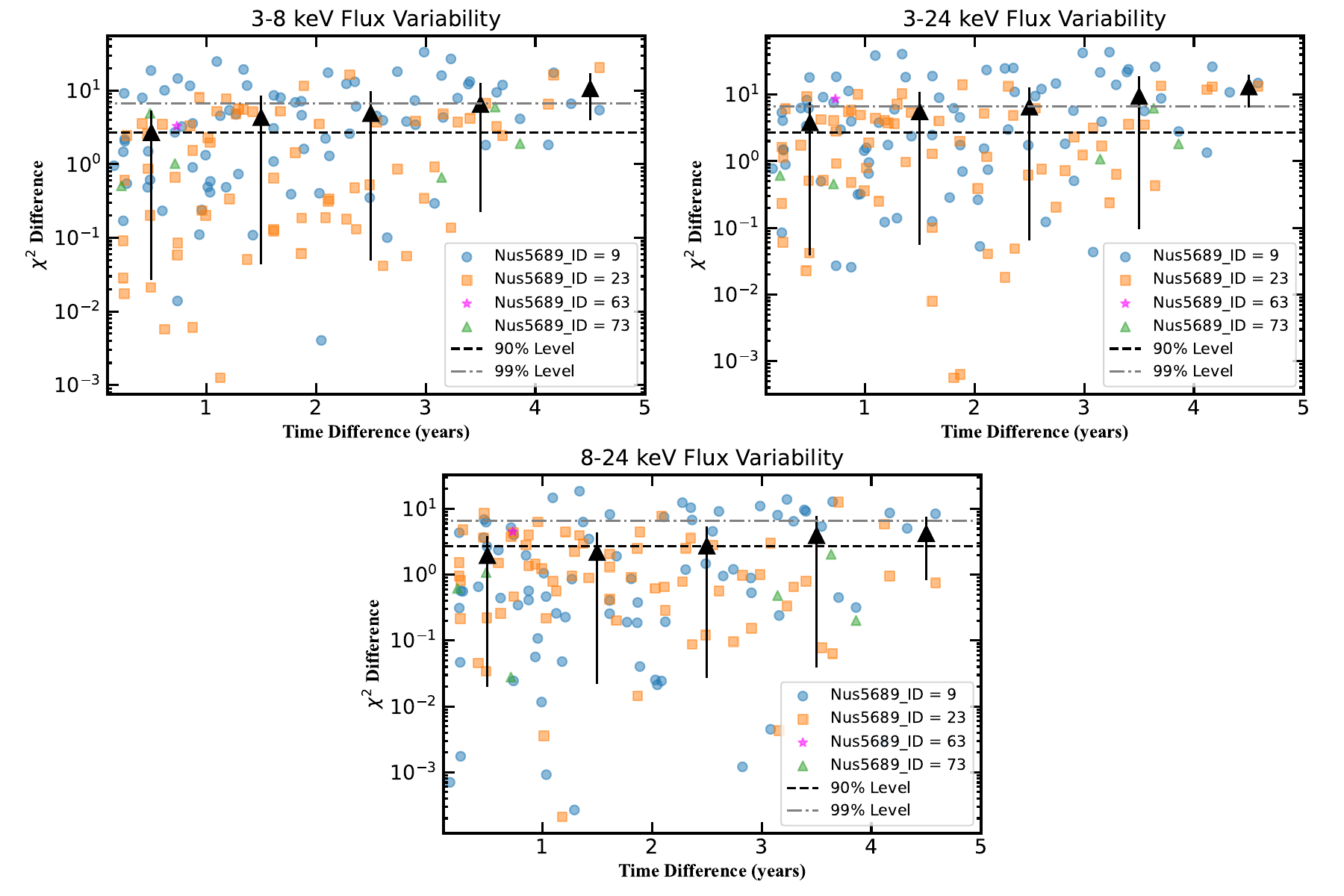}
    \caption{Structure function for every pair of observations for the four variable \nus sources. Point shapes and colors identify the sources as indicated in the legend. The three panels show the three energy bands as labeled. In all three plots, the black triangles are the averages in each time bin (5 bins, 1 year in width), and the error bars are the standard deviation (1$\sigma$) in that bin. Above the black dotted lines represent where the observations vary more than a 90\% confidence level, and above the grey dash-dotted line is where they vary above a 99\% confidence level.}
    \label{fig:nus_time_scale_var}
\end{figure*}

Figure \ref{fig:nus_time_scale_var} shows the $\chi^2$ for all observation pairs for the four variable sources, i.e., the structure function. Table \ref{tab:time_scale_fract} shows the fraction of observations that exhibit variability at the 90\% and 99\% confidence levels for each time bin and energy band. For all three bands, the 4--5 year bin has a significantly higher fraction of variable observation pairs compared to the 0--1 year bin. This difference ranges from a factor of $\sim$2 up to $\sim$5. While there is variability on all timescales, it is most likely on timescales $\gtrsim$4 years.
This behavior is consistent with ``red-noise variability'', i.e. when AGN exhibit larger-amplitude variability on longer timescales \citep{Lawrence1987, Padovani2017, Chen2022}.

\renewcommand*{\arraystretch}{1.1}
\begin{table*}
    \caption{Variability Fraction Versus Time Interval}
    \centering
    \label{tab:time_scale_fract}
    \begin{tabular}{cccccccc}
    \hline\hline
    \textbf{Telescope} & \textbf{Band} & \textbf{CL} & \textbf{Year 0--1} & \textbf{Year 1--2} & \textbf{Year 2--3} & \textbf{Year 3--4} & \textbf{Year 4--5}\\
    \hline
    \nus &  \thei & 90\% & 0.31 & 0.50 & 0.45 & 0.68 & 0.86 \\
    & & 99\% & 0.14 & 0.23 & 0.21 & 0.36 & 0.43 \\ 
    & \thtw & 90\% & 0.49 & 0.45 & 0.42 & 0.68 & 0.86 \\
    & & 99\% & 0.18 & 0.20 & 0.27 & 0.36 & 0.86 \\ 
    & \eitw & 90\% & 0.31 & 0.23 & 0.30 & 0.41 & 0.57 \\
    & & 99\% & 0.04 & 0.08 & 0.21 & 0.27 & 0.29 \\ 
    \hline
    \xmm & \zetw & 90\% & 0.34 & 0.40 & 0.53 & 0.55 & \no \\
    & & 99\% & 0.14 & 0.28 & 0.30 & 0.39 & \no \\ 
    & \twte & 90\% & 0.34 & 0.39 & 0.53 & 0.54 & \no \\
    & & 99\% & 0.14 & 0.27 & 0.30 & 0.38 & \no \\ 
    \hline
    \end{tabular}
\raggedright
\tablecomments{Each number in the five rightmost columns is the fraction of observation pairs for the 4 variable \nus sources and 11 variable \xmm sources that display variability in the $\Delta t$ bin indicated at top and the observatory, energy band, and confidence level indicated at the left of each row.}
\end{table*}

\begin{figure*}
    \centering
    \includegraphics[scale=0.75]{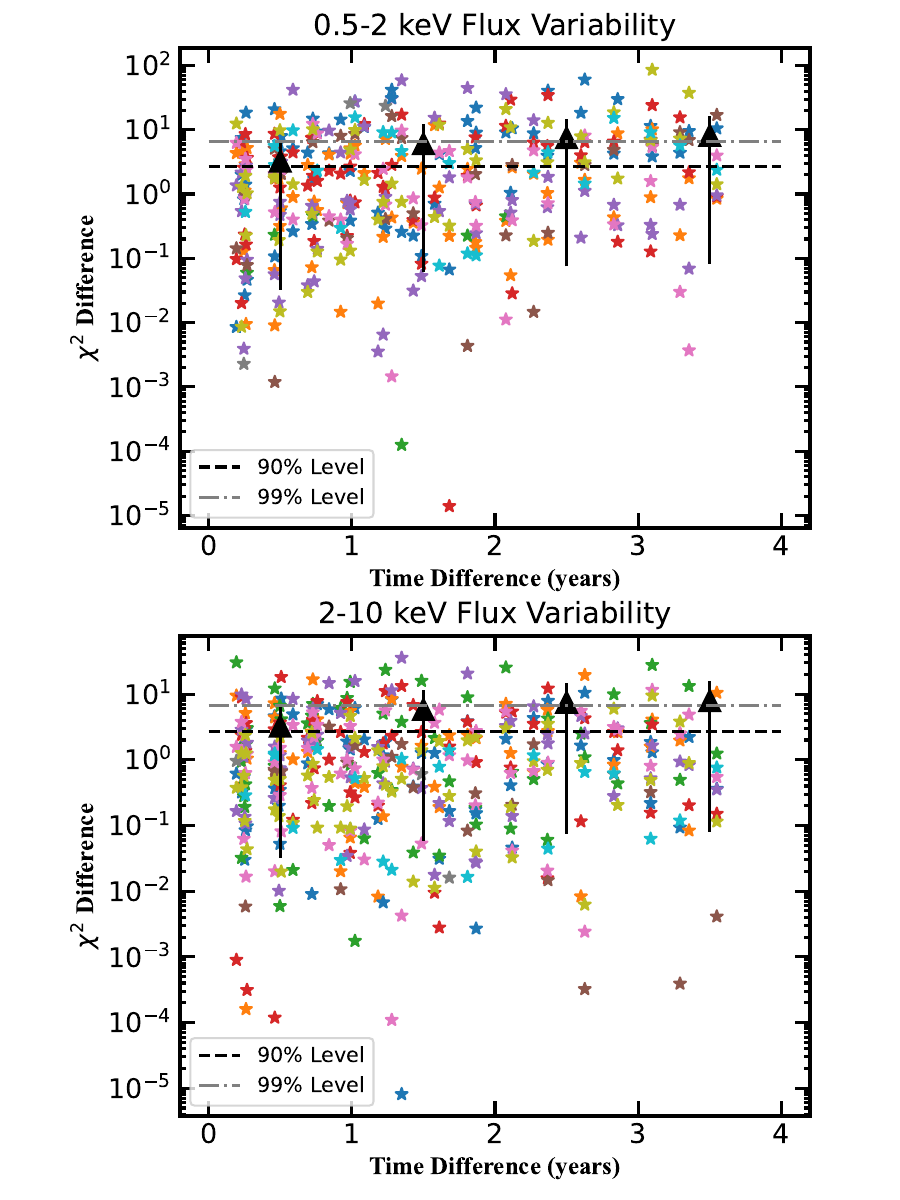} 
    \caption{Structure function for every pair of observations for the 11 variable \xmm sources. The top plot shows the results for the \zetw band, and the bottom plot shows the results for the \twte band. In both plots, the black triangles are the averages in each time bin (4 bins, 1 year in width) and the error bars are the standard deviation (1$\sigma$) in that bin. Above the black dotted lines represent where the observations do not agree within a 90\% confidence level, and above the grey dash-dotted line is where the observations do not agree within a 99\% confidence level.}
    \label{fig:xmm_time_scale}
\end{figure*}

XMM observations of the NEP-TDF did not start until \nus cycle 6 (2020 October). That makes the maximum time between observations just over 3.5 years, and the minimum time is 3 months. 
Figure \ref{fig:xmm_time_scale} shows the XMM structure functions for the 11 variable sources. The fractions of observation pairs showing variability at the 90\% and 99\%  levels for the \zetw and \twte bands are listed in Table \ref{tab:time_scale_fract}. While not quite as strong as the \nus results, there is still a systematic increase in the variability fraction in the 3--4 year bin compared to the 0--1 bin year. However, the small sample sizes make these results uncertain.
A followup paper (S.\ Creech et~al., ApJ, submitted, 2026) will provide further  analysis of variability across different energy bands and timescales.

\subsection{Comparison with Previous Surveys}

\subsubsection{Variability Fraction}

Previous X-ray variability studies include several based on \cha data in the CDF-S\null. The earliest \citep{Paolillo2004} analyzed $\sim$1\,Ms of data and found 74 of 340 sources ($\sim$20\%) displayed variability. However, when only sources with $>$100 counts were considered, this fraction increased to 45\%, and the variable fraction increased to 94\% for sources with $>$800 counts. Later, \cite{Young2012} analyzed  4\,Ms of CDF-S data and found similar results: of the 369 AGN analyzed, 185 (50\%) were variable, and for sources with $\gtrsim$800 counts, the variable fraction was $\sim$80\%. Finally, \cite{Yang2016} used 6\,Ms of observations to look for variability of the 68 brightest radio-quiet AGN\null. These sources all had $>$500 counts, and 61 (90\%) displayed significant variability.

In other fields, \cite{Bauer2004} investigated the 136 AGN detected in the \cha 2\,Ms CDF-N survey and found 81 (60\%) were variable. Moreover, in the sample of 61 sources with $>$500 counts, this fraction increased to $\sim$80--90\%. 
Finally, \cite{Lanzuisi2014} used the $\sim$1.55\,Ms of the XMM-COSMOS survey to analyze variability in the 638 AGN detected. Of these 638, 200 (31\%) exhibited significant flux variability after allowing for false positives. As for the previous surveys mentioned, limiting the sample to sources with $>$1000 counts increased the variable fraction to 75\%. 

The present work finds a smaller percentage of variable sources: only 2\% of XMM sources and 4\% of \nus sources exhibited variability. However, this is due to the limited statistics of this survey: the total XMM exposure time of the NEP-TDF survey was at least 4$\times$ less than the other three surveys. We had far less XMM exposure time because our survey is focused on the hard X-rays, while the other surveys are exclusively soft X-ray surveys. However, for the brightest sources ($>$600 XMM counts), 5/7 (71\%) sources are variable. Additionally, one of the two non-variable sources, XMM69 108, just missed the cutoff with $p = 0.007$. Moreover, of the 3 \nus sources with at least 600 counts in the \thtw band, 2 exhibit variability (66\%). Therefore, our results are consistent with previous X-ray surveys, all of which point to the same conclusion: the vast majority of AGN ($\gtrsim$ 70\%) are variable when the source statistics are sufficient to detect it. 

\subsubsection{Variability Timescales}

In their sample of 136 AGN in the CDF-N, \cite{Bauer2004} found that 60\% of sources displayed variability on day-to-year timescales versus only 55\% on second-to-day timescales. This fraction increased to 90\% for day-to-year timescales versus 80\% for second-to-day for the 61 sources with $>$500 counts. 
\cite{Paolillo2004} found corroborating results in the first 1\,Ms of the CDF-S. While variability in $<$2 days was often detected, their sample was dominated by long-term variability. In fact, detecting variability on timescales longer than one month was twice as likely as detecting it in observations separated by less than 10 days. 

\cite{Yang2016} had the longest time separation of all the works mentioned. In 15 years of monitoring, 90\% of their sample of 68 sources displayed significant variability between the first and last observation. Taking all of these results in conjunction with those presented in Section~\ref{sec:var_time}, the data support AGN being more variable on month-to-year timescales  than on day-to-month timescales. More precise measurements will require more complex statistical techniques that take into account the measurements for sources not individually detected as variable.

\section{Summary and Conclusions} \label{sec:sum+concl}

The key results from the present variability analysis are:

\begin{itemize}
    \item Of the 112 \nus sources,  4  displayed variability with $p \leq 0.01$ in at least one of the 3--24\,keV, 3--8\,keV, and 8--24\,keV bands. NuS59 9 was the only source that varied in all three bands,  and it exhibited the highest overall level of \nus variability ($p = 7 \times 10^{-16}$ in the \thtw band). 

    \item Of the 453 XMM sources, 11 were variable with $p\leq 0.002$ in at least one of the \zetw and \twte bands. Two sources exhibited variability in both bands: XMM69 299/305.
    
\item In the combined \nus and \xmm samples, 9/29 of the sources with $>$400 counts from at least one observatory were detected as variable.  Among fainter sources, only 4/475 showed detectable variability. Many sources not individually detected as variables probably vary at some level, and advanced statistical techniques could be able to measure the overall variability of the population.

   \item Two XMM sources, XMM69 137/327, match HST-variable sources \citep{Obrien2024}, but only XMM69 327 showed  X-ray variability ($p = 0.0009$ in the \zetw band).

    \item For the 3 \nus and 10 \xmm variable sources with known redshifts, the majority are at least 5$\times$ brighter than the survey limit. Moreover, 27\% of all sources above this 5$\times$ threshold are variable, while only 4\% of sources below it are variable. 

    \item Host galaxies of all variable sources with available stellar masses (R.\ Ortiz et~al., ApJ, submitted 2026) have $\log (M_*/M_\sun)>10.5$, while only 57\% of hosts of non-variable sources are this massive.  The association of higher host mass with variability appears real despite selection effects.  In contrast, host-galaxy SFR and AGN fraction showed no significant differences between variable and non-variable sources.    
    
    \item When enough counts are available, variability is near-ubiquitous in AGN, also in agreement with previous X-ray surveys.
    
    \item In both \xmm and all three \nus bands, variability is more likely between observations with at least 4 years of separation than those with less than one year of separation. In other words, the structure function shows a red-noise character. This agrees with results of previous X-ray surveys, and advanced statistical tests might be able to measure the population structure function.
\end{itemize}

Further analysis of AGN X-ray variability, and in particular an assessment of whether it comes from intrinsic luminosity changes or changes in obscuration, will be presented by S.\ Creech et~al.\ (ApJ submitted, 2026).

\begin{acknowledgments}
The material is based upon work supported by NASA under award number 80GSFC24M0006.\\
We appreciate the helpful comments from the referree that improved the quality of this paper.
\end{acknowledgments}

\restartappendixnumbering
\appendix

\section{Light Curves}

\begin{figure*}
    \centering
    \includegraphics[scale=0.5]{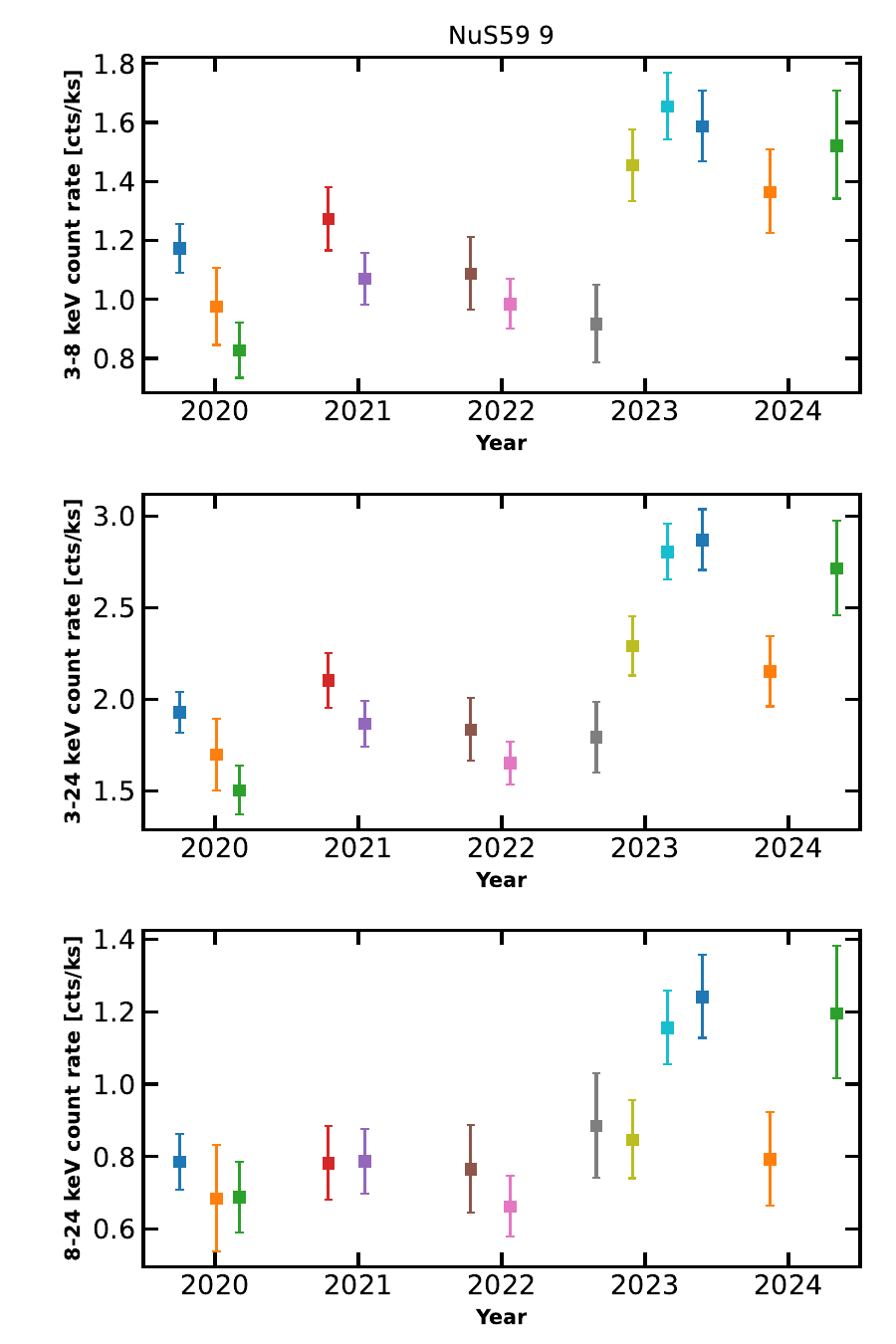} 
    \includegraphics[scale=0.5]{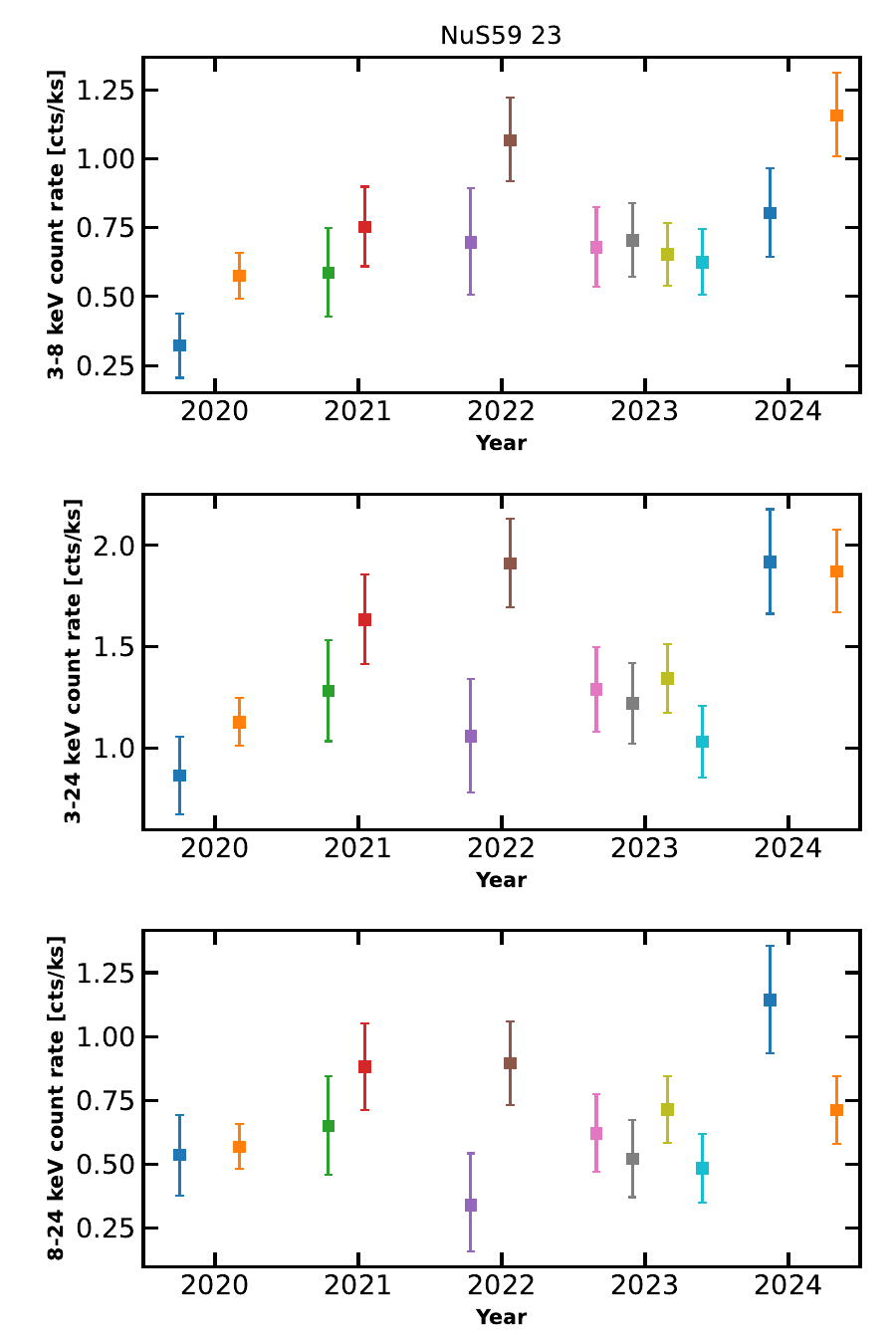}\\
    \includegraphics[scale=0.5]{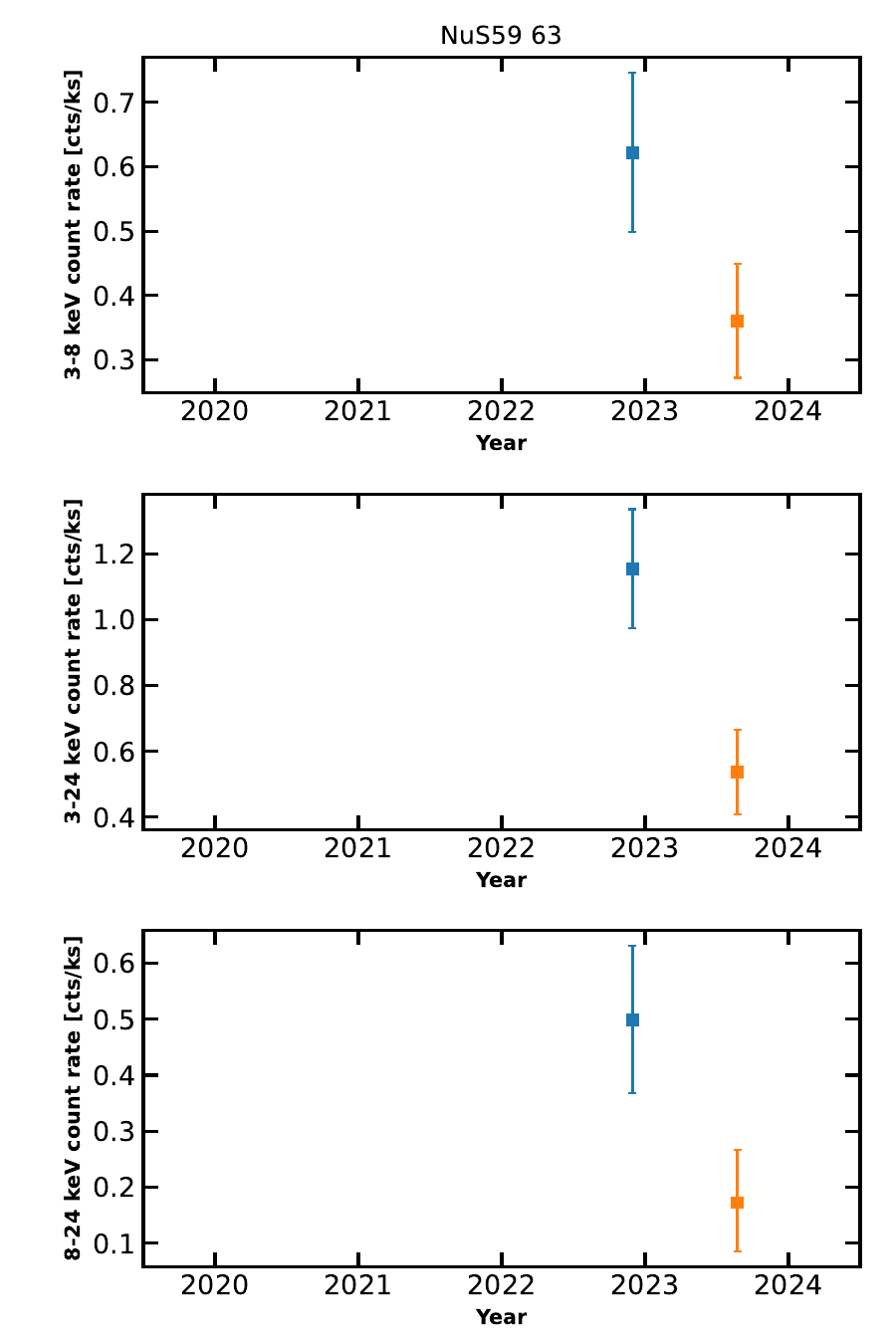} 
    \includegraphics[scale=0.5]{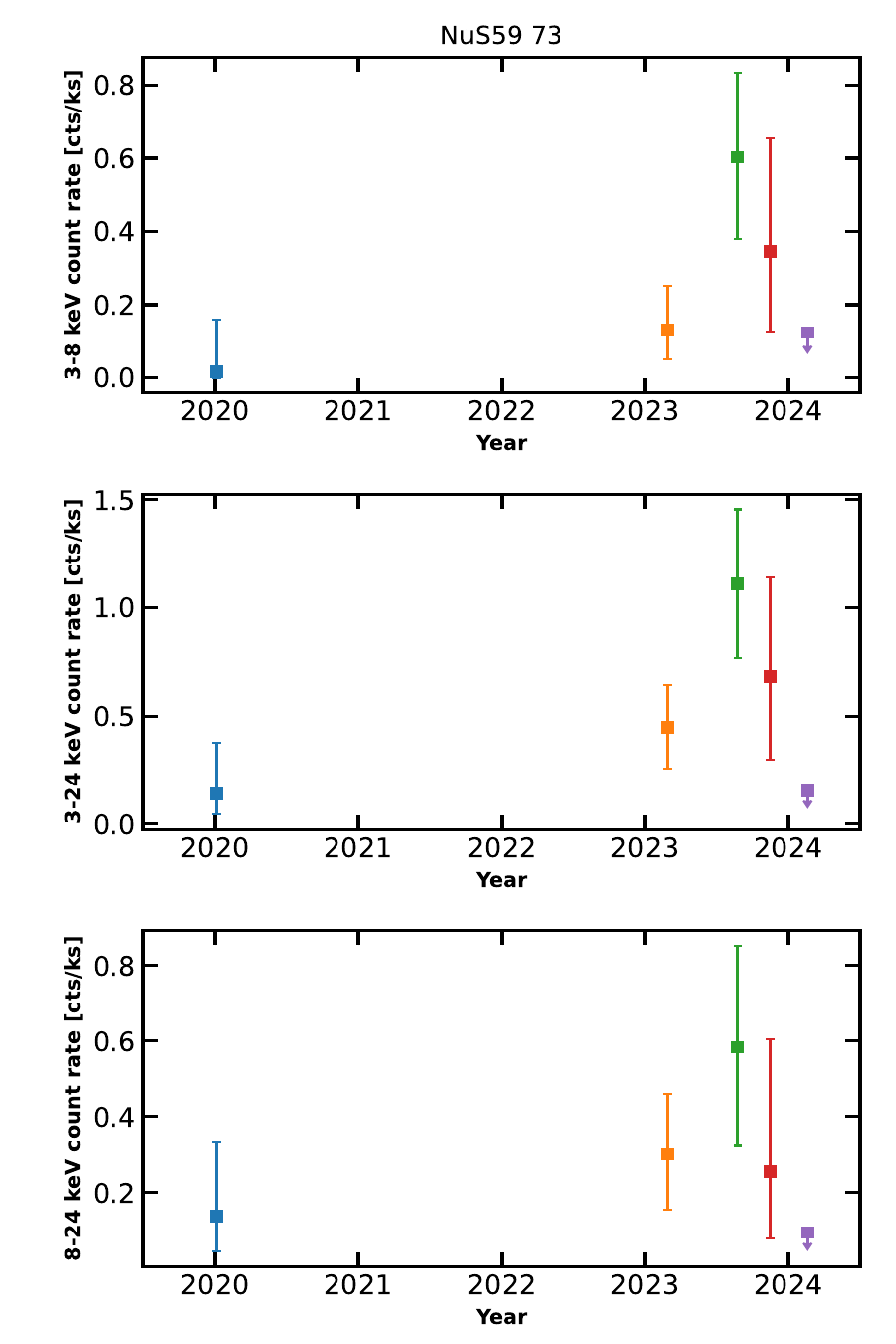}
    \caption{Count-rate light curves for the four \nus variable sources as labeled above each set of three panels. The three panels in each set show the three bands as labeled on each ordinate. Each color represents a different epoch.}
    \label{fig:lc_blazar}
    \label{fig:lc_nus_63-73}
\end{figure*}

\begin{figure}
    \centering
    \includegraphics[scale=0.5]{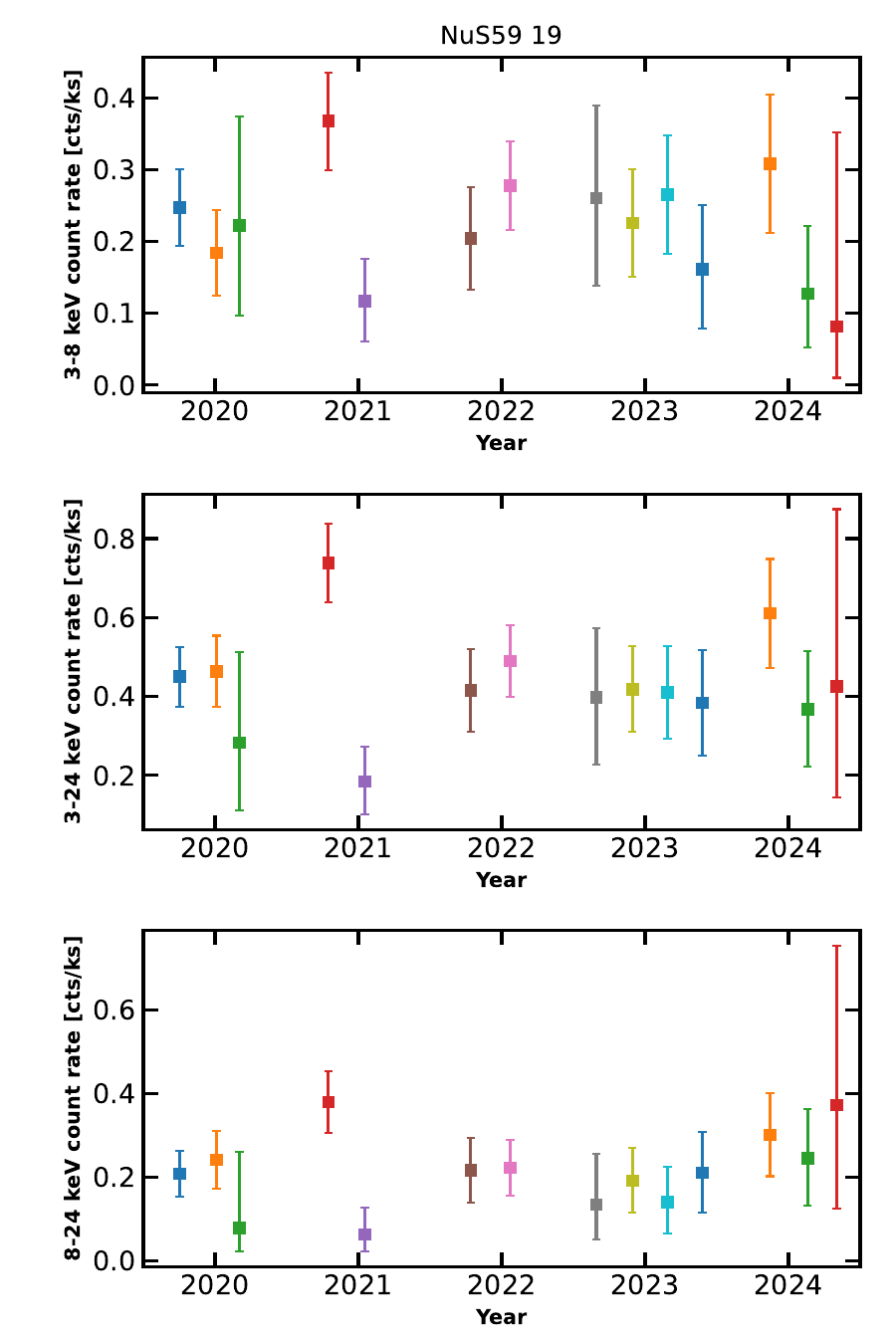} 
    \caption{Count-rate light curves for NuS59 19. The three panels show the three bands as labeled on each ordinate. Each color represents a different epoch.  \citetalias{Zhao2024} found this source to be variable with $p\leq 0.01$, but the source no longer meets this threshold after including the subsequent \nus observations.}
    \label{fig:lc_nus_19}
\end{figure}

Figure \ref{fig:lc_blazar} shows the light curves for the four variable \nus sources, and Figure \ref{fig:lc_nus_19} shows the light curves for the one source found to be variable by \citetalias{Zhao2024} but not in this work.
Figure~\ref{fig:lc_xmm_111-318}  shows the light curves for the 11 variable \xmm sources.

\begin{figure*}
    \includegraphics[scale=0.5, trim=0 80 0 20, clip]{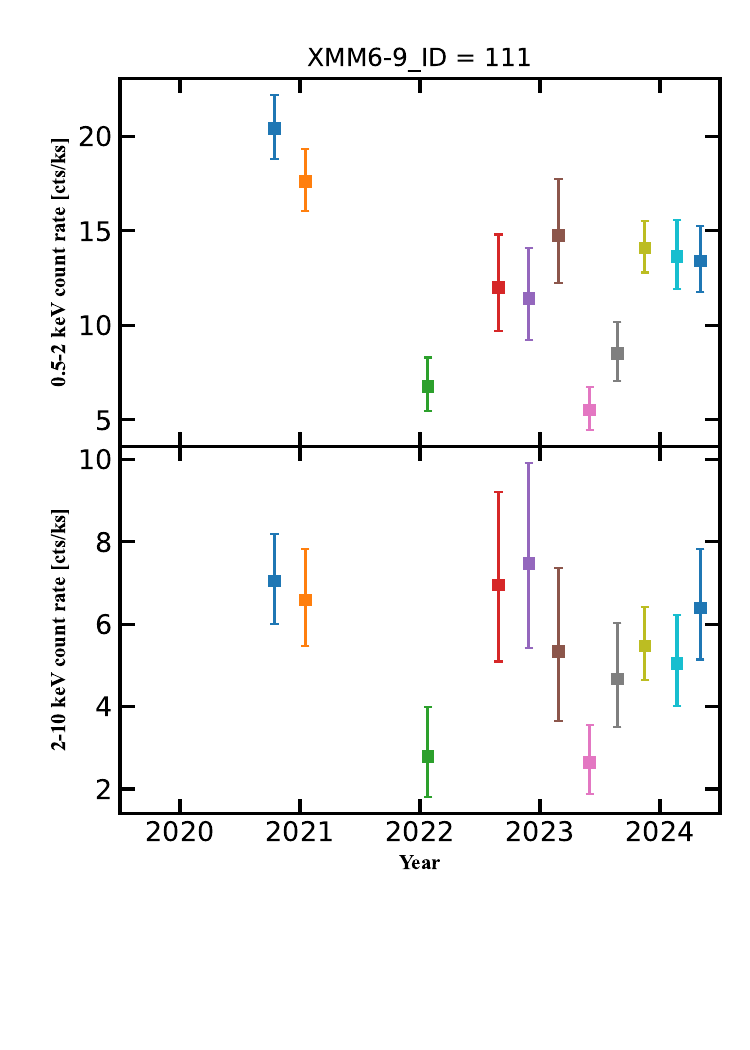} 
    \includegraphics[scale=0.5, trim=0 80 0 20, clip]{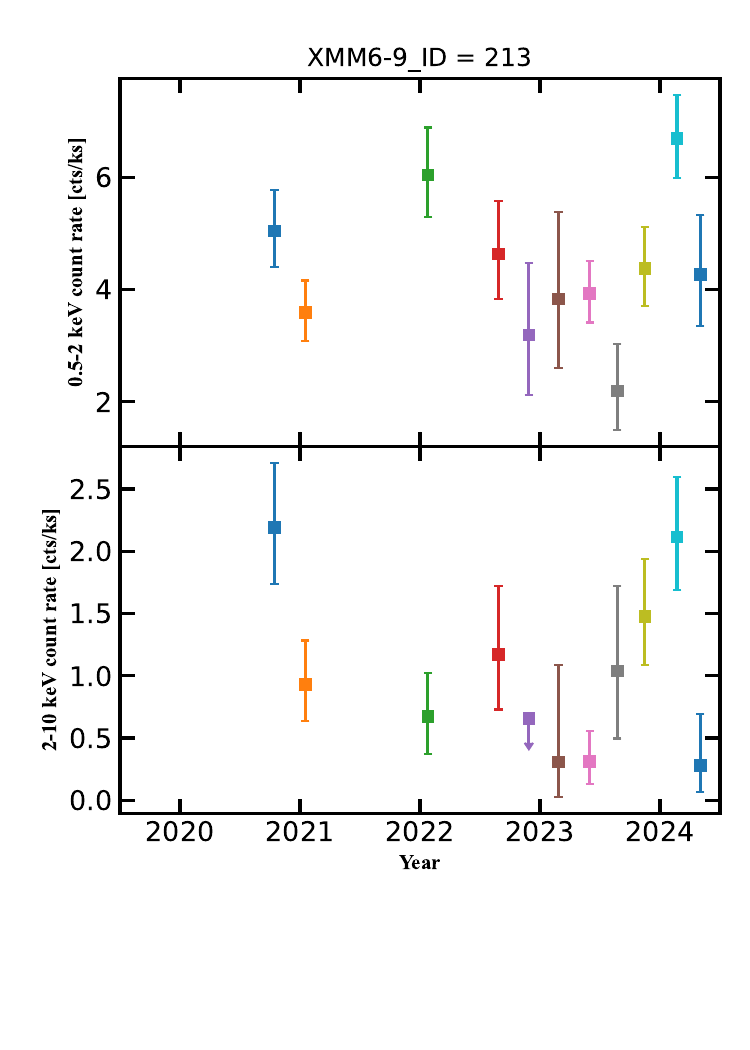}
    \includegraphics[scale=0.5, trim=0 80 0 20, clip]{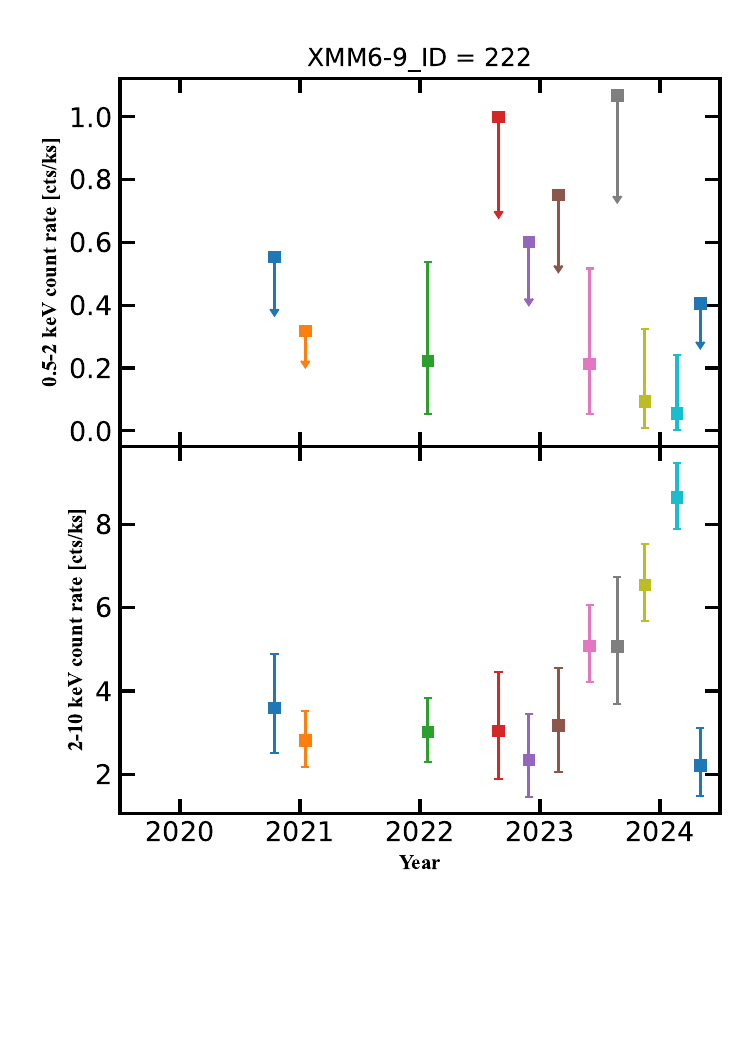} 
    \includegraphics[scale=0.5, trim=0 80 0 20, clip]{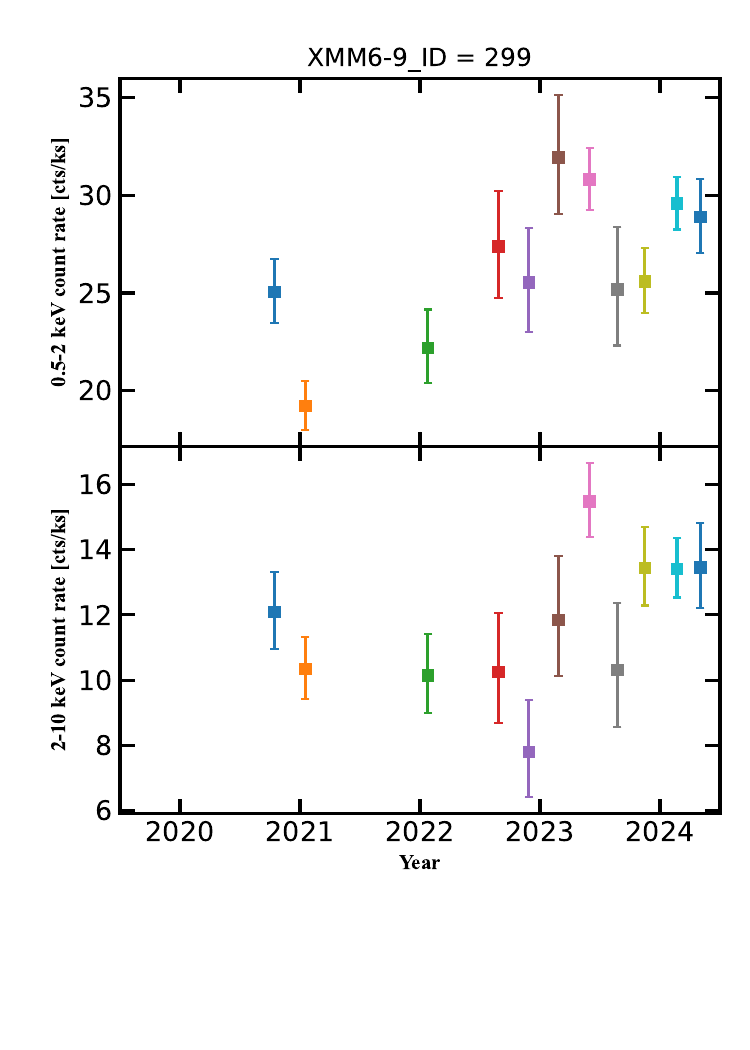}
    \includegraphics[scale=0.5, trim=0 80 0 20, clip]{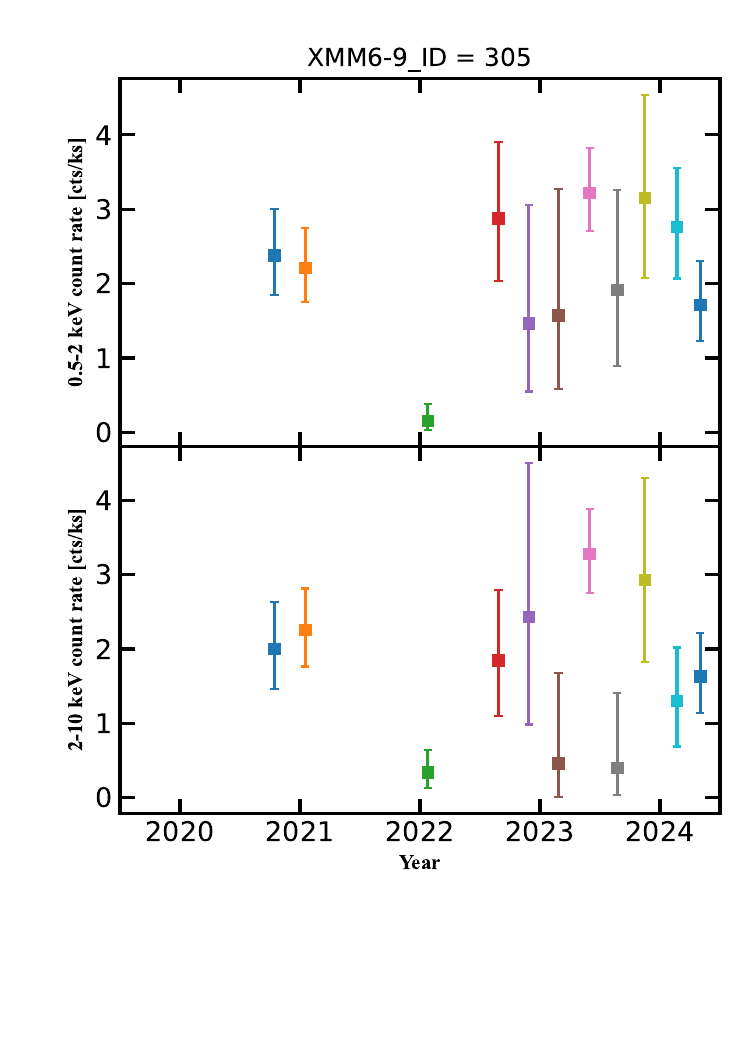} 
    \includegraphics[scale=0.5, trim=0 80 0 20, clip]{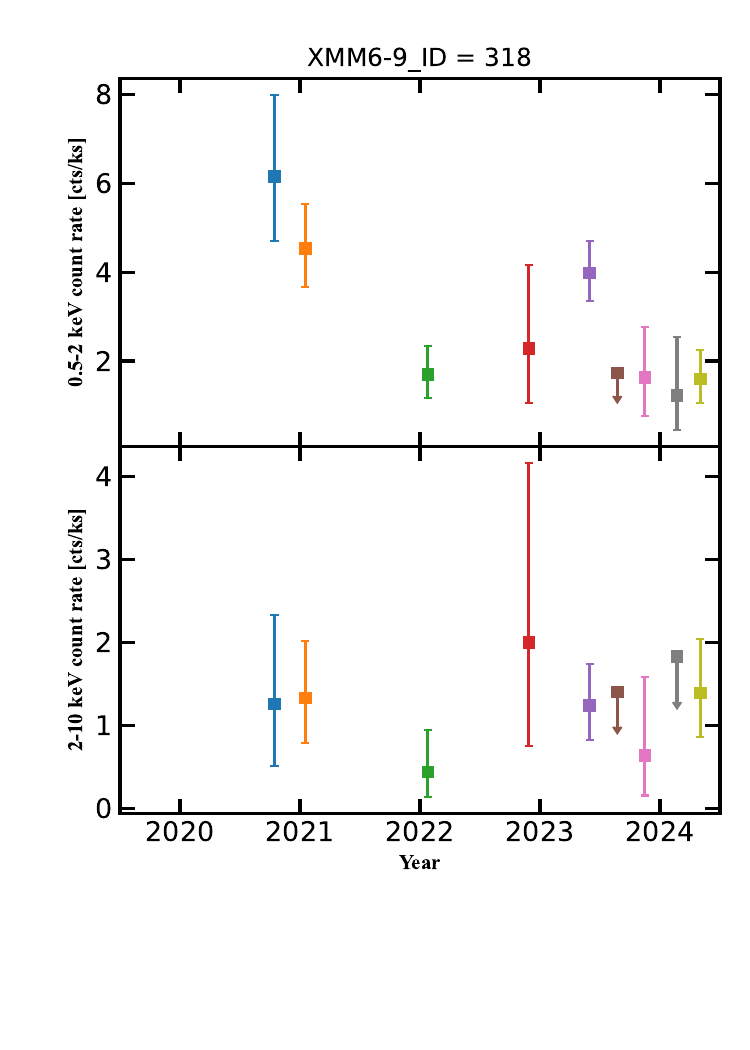}
    \caption{Count-rate light curves for the 11 variable \xmm sources as labeled above each pair of panels.
    The two panels in each set show the soft (top) and hard (bottom) bands  as labeled on each ordinate.
    Source IDs are labeled above each panel pair, and each color represents a different epoch.}
    \label{fig:lc_xmm_111-318}
\end{figure*}

\addtocounter{figure}{-1}
\begin{figure*}
    \includegraphics[scale=0.5, trim=0 80 0 20, clip]{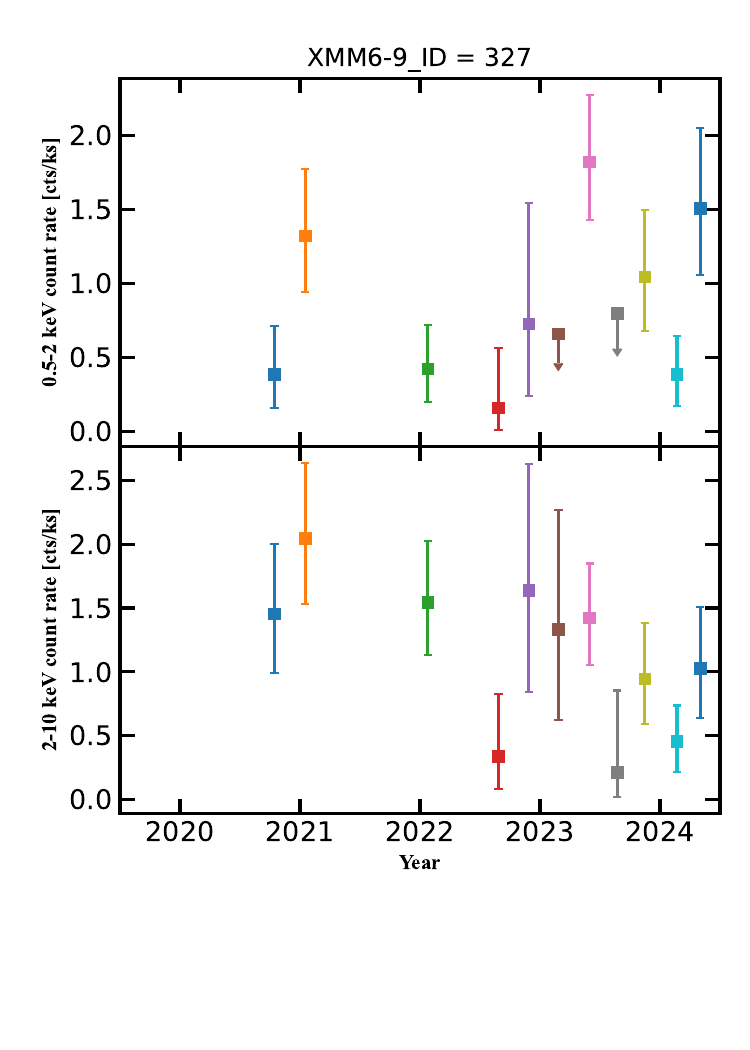} 
    \includegraphics[scale=0.5, trim=0 80 0 20, clip]{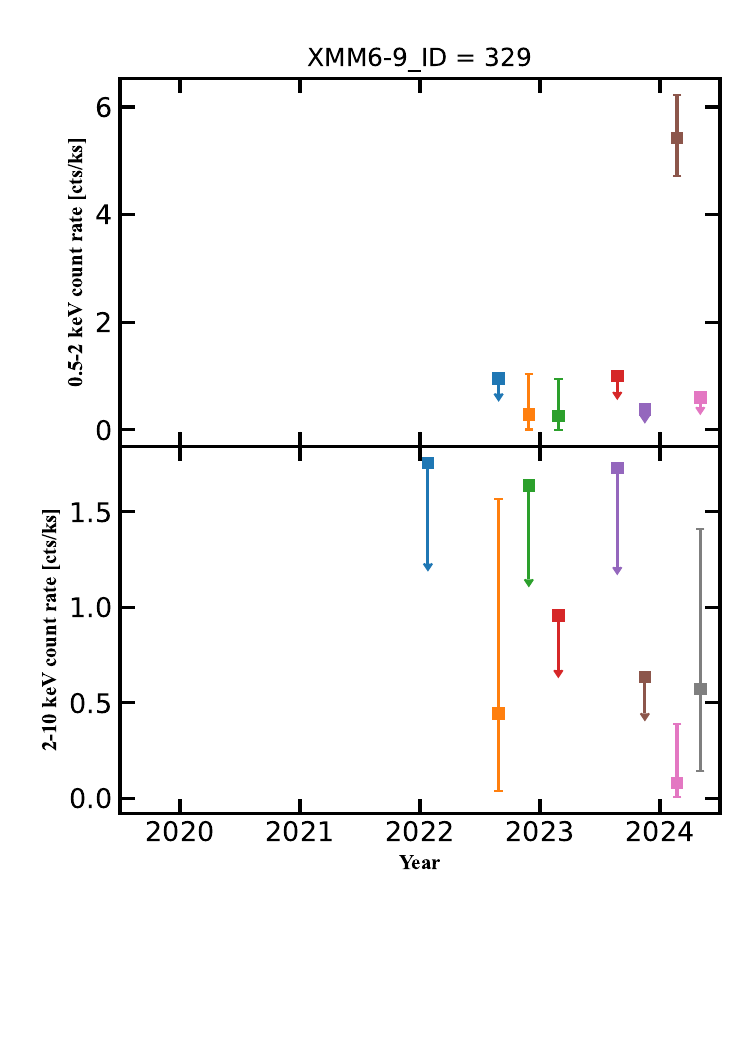}
    \includegraphics[scale=0.5, trim=0 80 0 20, clip]{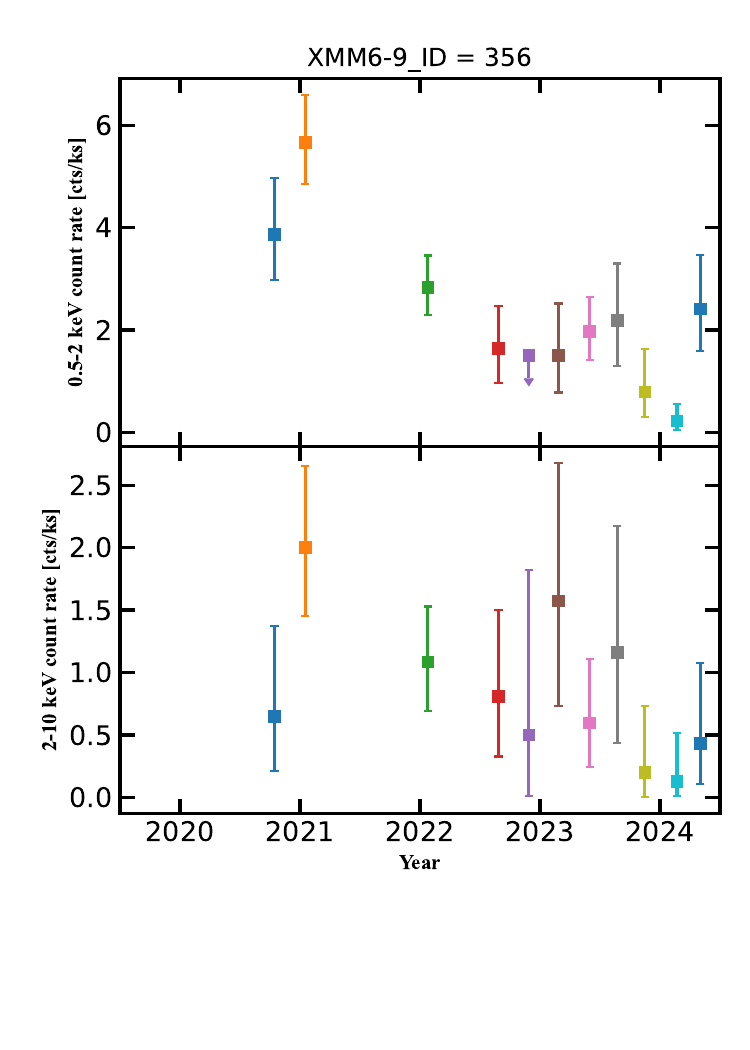} 
    \includegraphics[scale=0.5, trim=0 80 0 20, clip]{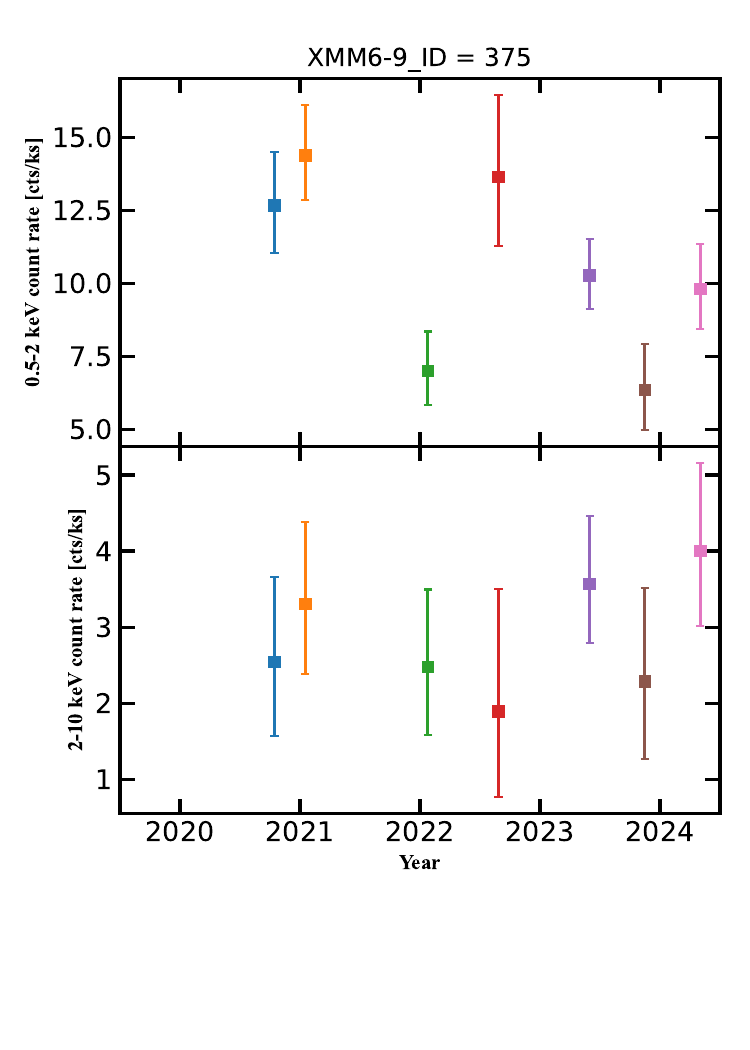}
    \includegraphics[scale=0.5, trim=0 80 0 20, clip]{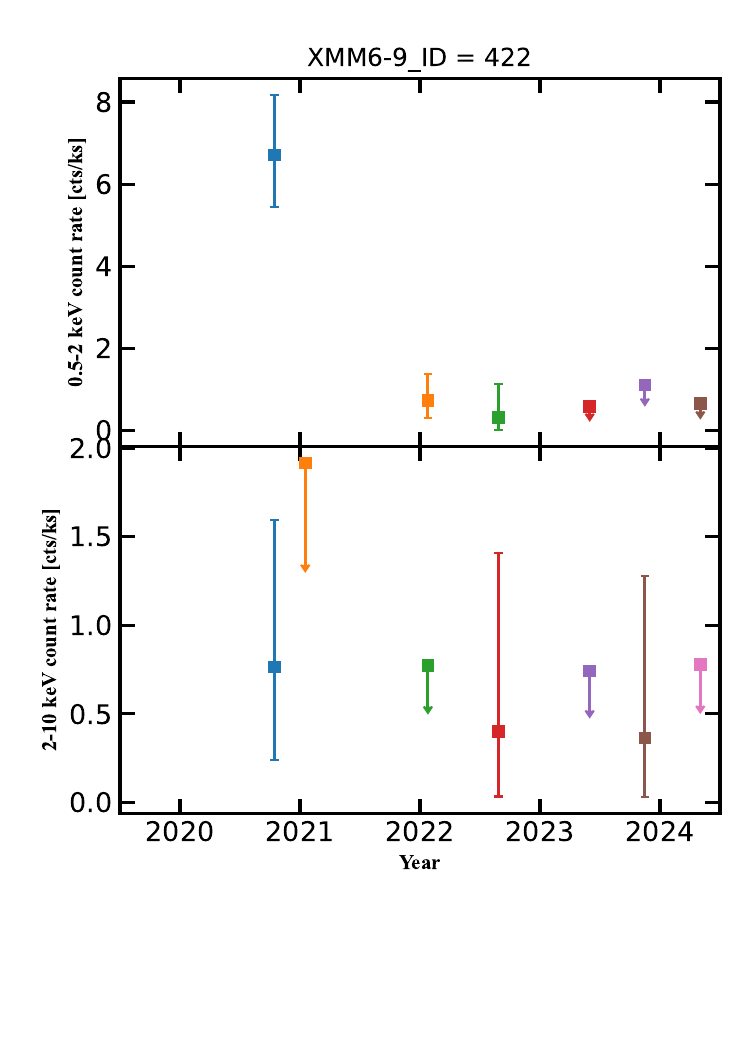}
    \caption{Continued}
    \label{fig:lc_xmm_327-422}
\end{figure*}

\section{Catalog Descriptions} \label{app_cat_desc}
The catalogs are available in the source directory with the \nus catalog  as file ``nustar\_catalog.csv'' and the \xmm  catalog as file ``xmm\_catalog.csv''.

\FloatBarrier
\renewcommand{\arraystretch}{1.05}
\begin{table*}[b!]
\caption{\nus Catalog Description}
\centering
\label{Table:nus_cat_info}
  \begin{tabular}{ll}
   \hline
   \hline
 	Col. & Description \\
	\hline
    1 & Nus59 ID number \\
    2 & \nus source name (use “NuSTAR JHHMMSS+DDMM.m”) \\
    3 & Nus56 ID number \citep{Zhao2024} \\
    4 & Nus89 ID number \citep{Silver2026} \\
    5 & XMM69 ID number (Table~\ref{Table:xmm_cat_info}) \\
    6--7 & \nus coordinates (J2000) of the source in whichever energy band has the highest DET\_ML \\
    8 & Spectroscopic redshift of the associated source \\
    9 & Photometric redshift of the associated source \\
    10 & Spectroscopic classification (Q for quasars, G for galaxies, S for stars, N if no measurement) \\
    11 & Variability $p$ in the \thei band \\
    12 & Variability $p$ in the \eitw band \\
    13 & Variability $p$ in the \thtw band \\
    14 & log Stellar Masses ($M_{\sun}$) \\
    15 & log Star-Formation Rate ($M_{\sun}$ yr$^{-1}$) \\
    16 & total \nus counts in all epochs, all bands\\   \hline   	
\end{tabular}
\raggedright
\end{table*}

\begin{table*}[h!]
\caption{\xmm  Catalog Description}
\centering
\label{Table:xmm_cat_info}
  \begin{tabular}{ll}
   \hline
   \hline
 	Col. & Description \\
	\hline
    1 & XMM69 ID number \\
    2 & XMM6  ID number \citep{Zhao2024}\\
    3 & XMM89  ID number \citep{Silver2026}\\
    4 & \xmm source name (use “TDFXMM JHHMMSS+DDMM.m”) \\
    5 & Nus59 ID number (Table~\ref{Table:nus_cat_info}) \\
    6--7 & X-ray coordinates (J2000) of the source in whichever energy band has the highest DET\_ML \\
    8 & Spectroscopic redshift of the associated source \\
    9 & Photometric redshift of the associated source \\
    10 & Spectroscopic classification (Q for quasars, G for galaxies, S for stars, N/A if no measurement) \\
    11 & Variability $p$ in the \zetw band \\
    12 & Variability $p$ in the \twte band \\
    13 & total \xmm counts in all epochs, both bands\\
   \hline   	
\end{tabular}
\raggedright
\end{table*}

\bibliographystyle{aasjournal}
\bibliography{bibliography}

\end{document}